\documentclass[a4paper,11pt]{article}
\usepackage{jheppub} 
\usepackage[T1]{fontenc} 

\usepackage{graphicx,multirow,subfigure}
\usepackage{float}
\usepackage{bm}
\usepackage{amsmath}
\usepackage{amssymb}
\usepackage{amscd}
\usepackage{latexsym}
\usepackage{slashed}
\usepackage{color}
\usepackage{graphicx}
\usepackage[normalem]{ulem}
\usepackage{color}
\definecolor{orange}{cmyk}{0,0.5,1,0}
\usepackage{verbatim}
\usepackage{rotating}
\usepackage{diagbox}
\usepackage{cancel}
\usepackage[utf8]{inputenc}

\def\be{\begin{equation}}
\def\ee{\end{equation}}
\def\bea{\begin{eqnarray}}
\def\eea{\end{eqnarray}}
\def\nn{\nonumber}

\title{Naturalness and Dark Matter in the  BLSSM }

\author[a,b]{Luigi Delle Rose,}
\author[c]{Shaaban Khalil,}
\author[a]{Simon J.D. King,}
\author[d]{Carlo Marzo,}
\author[a,b]{Stefano Moretti}
\author[e]{and Cem S. Un}

\emailAdd{L.Delle-Rose@soton.ac.uk}
\emailAdd{Skhalil@zewailcity.edu.eg}
\emailAdd{SJD.King@soton.ac.uk}
\emailAdd{Carlo.Marzo@kbfi.ee}
\emailAdd{S.Moretti@soton.ac.uk}
\emailAdd{cemsalihun@uludag.edu.tr}

\affiliation[a]{School of Physics and Astronomy, University of Southampton, Highfield, Southampton SO17 1BJ, United Kingdom}
\affiliation[b]{Particle Physics Department, Rutherford Appleton Laboratory, Chilton, Didcot, Oxon OX11 0QX, United Kingdom}
\affiliation[c]{Center for Fundamental Physics, Zewail City of Science and Technology, Sheikh Zayed,12588 Giza, Egypt}
\affiliation[d]{National Institute of Chemical Physics and Biophysics, R{\"a}vala 10, 10143 Tallinn, Estonia}
\affiliation[e]{Department of Physics, Uluda\~{g} University, TR16059 Bursa, Turkey}

\abstract{We study the naturalness properties of the $B-L$ Supersymmetric Standard Model (BLSSM) and compare them to
those of the Minimal Supersymmetric Standard Model (MSSM) at both low (i.e.,  Large Hadron Collider) energies
and high (i.e., unification) scales. By adopting standard measures of naturalness, we assess that, in presence of full unification of the additional gauge couplings and scalar/fermionic masses of the BLSSM, such a scenario reveals a somewhat 
higher degree of Fine-Tuning (FT) than the MSSM, when the latter is computed at the unification scale and all available theoretical and experimental constraints, but the Dark Matter (DM) ones, are taken into account. Yet, such a difference, driven primarily by the collider limits  requiring a high mass for the gauge boson
associated to the breaking of the additional $U(1)_{B-L}$ gauge group of the BLSSM in addition to the 
$SU(3)_C\times SU(2)_L \times U(1)_Y$ of the MSSM, 
 should be regarded as a modest price to pay for the former  in relation to the latter, if
one notices 
that the non-minimal scenario offers a significant volume of parameter space where numerous DM solutions of different compositions
can be found to the relic density constraints, unlike the case of the minimal structure, wherein only one type of solution is
accessible over an ever diminishing parameter space. In fact, this different level of tension within the two SUSY models
 in complying with current data is 
well revealed when the FT measure is recomputed in terms of the low energy spectra of the two models, over their allowed regions of parameter space now in presence of all DM bounds, as it is shown that the tendency is now opposite, the BLSSM appearing more natural than the MSSM.}

\begin{document} 
\maketitle
\flushbottom

\section{Introduction}
\label{sec:introduction}

Low scale Supersymmetry (SUSY) is motivated by solving two major flaws  of the Standard Model (SM):  the gauge hierarchy  and  Dark Matter (DM) problems. 
In the SM, the hierarchy problem stems from the fact that a very unnatural Fine-Tuning (FT) is required to keep the Higgs mass at an acceptable value for current data. SUSY provides an elegant solution to this. However, SUSY must be broken at a high scale, hence some FT is reintroduced at  some level. In the Minimal Supersymmetric Standard Model (MSSM), with universal soft SUSY breaking terms, a heavy spectrum is required to give large radiative corrections to the SM-like Higgs mass and account for the recently measured value of 125 GeV at the Large Hadron Collider (LHC). Thus naturalness becomes seriously challenged in the MSSM by well established experimental conditions. 

Also, the alluring hints of DM existence are serious indications for new physics Beyond the SM (BSM). Due to $R$-parity conservation, the Lightest SUSY Particle (LSP) in the MSSM, the lightest neutralino, is stable and thus is a good candidate for DM. However, the  constraints from LHC data (from the Higgs boson properties as well as  void
searches for additional Higgs and SUSY states)
combined with cosmological relic density and DM direct detection data rule out all of the MSSM
parameter space except a very narrow region of it \cite{Abdallah:2015hza}.

Quite apart from the aforementioned two problems of the SM, it should be recalled that non-vanishing neutrino masses are presently some of the most important evidence for BSM physics. Massive neutrinos are not present in the SM. However, a simple extension of it, based on the gauge
group $SU(3)_C \times SU(2)_L \times U(1)_Y  \times U(1)_{B-L}$, can account for current experimental results of light neutrino
masses and their large mixing \cite{Khalil:2006yi,Basso:2008iv,Basso:2009gg,Basso:2010yz,Basso:2010as,Majee:2010ar,Li:2010rb,Perez:2009mu,Emam:2007dy,Khalil:2012gs,Khalil:2013in}.  Within the $B-L$ Supersymmetric Standard Model (BLSSM),  the SUSY version of such a scenario, which inherits the same beneficial features of the MSSM in connection with SUSY dynamics, 
it has been emphasised that the scale of $B-L$ symmetry breaking is related to the SUSY breaking one and both 
 occur in the TeV region \cite{Khalil:2016lgy,Khalil:2007dr,FileviezPerez:2010ek,CamargoMolina:2012hv,Kikuchi:2008xu,Fonseca:2011vn}. Therefore, several testable signals of the BLSSM are predicted for the current experiments at the LHC \cite{Elsayed:2011de,Basso:2012tr,O'Leary:2011yq,Basso:2012ew,Elsayed:2012ec,Khalil:2015naa,Abdallah:2014fra,Basso:2012gz,Abdallah:2015hma,Abdallah:2015uba,Hammad:2016trm,Hammad:2015eca}. 

In addition, the BLSSM provides new candidates for DM different from those of the MSSM. In particular, there are two kinds of neutralinos, corresponding to the gaugino of $U(1)_{B-L}$ and the $B-L$ Higgsinos. Also a right-handed sneutrino, in a particular region of parameter space, may be a plausible candidate for DM.  
We also consider the scenario where the extra $B-L$ neutralinos can be cold DM states. We then examine the thermal relic abundance of these particles and discuss the constraints imposed on the BLSSM parameter space from the negative results of their direct detection. We argue that, unlike the MSSM, the BLSSM offers one with significant parameter space satisfying all available experimental constraints. This may be at the expense of high FT, if $Z'$ is quite heavy and soft SUSY breaking terms are universal. Nevertheless, for what we will eventually verify to be a small increase in FT with respect to the MSSM, we will gain in the BLSSM a more varied DM sector and much better compliance with relic and (in)direct detection data. 

In the build-up to this DM phenomenology, we analyse the naturalness 
 problem in the BLSSM and compare its performance in this respect against that of the MSSM. In the latter, the weak scale ($M_Z$) depends on the soft SUSY breaking terms through the Renormalisation Group Equations (RGEs) and the Electro-Weak (EW) minimisation conditions, which can be expressed as 
\bea
\frac{1}{2} M_Z^2= \frac{m_{H_d}^2 - m_{H_u}^2 \tan^2 \beta}{\tan^2 \beta -1} -  \mu^2  . 
\label{EWmin}
\eea
Therefore, a possible measure of FT is defined as \cite{Barbieri:1998uv}
\bea 
\Delta(M_Z^2, a) = \left \vert \frac{a}{M_Z^2} \frac{\partial M_Z^2}{\partial a} \right\vert, 
\eea
where $a$ stands for the Grand Unification Theory 
(GUT) scale parameters (e.g., $m_0, m_{1/2}, A_0$, etc.) or low scale parameters (e.g., $M_1,M_2,M_3, m_{\tilde{q}}, m_{\tilde{\ell}}$, etc.). In order for SUSY to stabilise the weak scale, $\Delta \equiv {\rm Max} \left( \Delta(M_Z^2, a)\right)$ should be less than ${\cal O}(100~{\rm GeV})$. However, as the scale of SUSY breaking is increased, the EW one becomes highly fine-tuned. As intimated, in the BLSSM, both the weak and $B-L$ scales are related to soft SUSY breaking terms and, in addition to Eq.~(\ref{EWmin}), which is slightly modified by the presence of the gauge mixing $\tilde g$, we also have, in the same limit $\tilde g \simeq 0$, 
\bea
\frac{1}{2} M_{Z^\prime}^2= \frac{m_{\eta_1}^2 \tan^2 \beta' - m_{\eta_2}^2 }{1-\tan^2 \beta'} -  \mu'~^2 ,
\label{BLmin}
\eea
where $\eta_{1,2}$ are scalar bosons, with $\langle \eta_{1,2} \rangle = v'_{1,2}$ that break the $B-L$ symmetry spontaneously, and $\tan \beta' = v'_1/v'_2$. 
The bound on $M_{Z'}$,  due to negative searches at LEP, is given by $M_{Z'} /g_{BL} > 6$ TeV \cite{Cacciapaglia:2006pk}.
Furthermore, LHC constraints from the Drell-Yan (DY) process also exist, which force the $B-L$ $Z'$ mass to be in the
 few TeV region. 
This indicates that $m_{\eta_{1,2}}$ and $\mu'$ are of order TeV. Therefore, in the scenario of universal soft SUSY breaking terms of the BLSSM, a heavy $M_{Z'}$ implies higher soft terms, hence the estimation of the FT is expected to be worse than in the MSSM. At this point, it is worth mentioning that the $Z'$ gauge boson in the BLSSM can have a large decay width, thus potentially evading LEP and LHC constraints, which are based on the assumption of a narrow decay width, hence on $Z'$ decays into SM particles and additional neutrinos only. While this has been proven to be possible in
a non-unified version of the BLSSM, wherein the aforementioned  limits can be relaxed and $M_{Z'}$ can be of order one TeV \cite{Abdallah:2015hma,Abdallah:2015uba}, it remains to be seen whether a similar phenomenology can occur in the unified version of it which we are going to deal with here.   

The paper is organised as follows. In Section 2 we briefly review the BLSSM  with
a particular emphasis on the $B-L$ minimisation conditions that relate the mass of the neutral gauge boson $Z'$ to the soft SUSY breaking terms and also the extended neutralino sector. Section 3 is devoted to study the RGEs of the BLSSM matter content as well as the gauge and Yukawa couplings. The collider and DM constraints are addressed in Section 4. 
 In Section 5 we investigate  the FT measures in the BLSSM versus the MSSM case. Section 6 presents our numerical results.  
Finally, our remarks and conclusions are given in Section 7.    

\section{The $B-L$ Supersymmetric Standard Model}
\label{sec:BLSSM}

In this section, we briefly review the BLSSM  with an emphasis on its salient features with respect to the MSSM. Even though its gauge group seems like a simple extension of the MSSM gauge group with a gauged $U(1)_{B-L}$ (hereafter, $B-L$ symmetry), it significantly enriches the particle content, which drastically changes the low scale phenomena. First of all, the anomaly cancellation in the BLSSM requires three  singlet fields and the right-handed neutrino fields are the most natural candidates to be included in the BLSSM framework. In this context, also the SUSY seesaw mechanisms, through which non-zero neutrino masses and mixing can be achieved consistently with  current experimental indications \cite{Wendell:2010md}, can be implemented. Besides,  $R-$parity, which is assumed in the MSSM to avoid fast proton decay, can be linked to the $U(1)_{B-L}$ gauge group and it can be preserved if the $B-L$ symmetry is broken spontaneously \cite{Aulakh:1999cd}, as is the case in the BLSSM studied here. 

Spontaneous breaking of the $B-L$ symmetry can be realised in a similar way to the Higgs mechanism. That is, one can introduce two scalar fields, denoted as $\eta_{1,2}$. These fields should carry non-zero $B-L$ charges to break the $B-L$ symmetry and they are preferably singlet under the MSSM gauge group so as not to spoil  EW Symmetry Breaking (EWSB). Thus, the Superpotential in the BLSSM can be written as 
\begin{eqnarray}
W &=&\mu H_{u}H_{d}+Y_{u}^{ij}Q_{i}H_{u}u^{c}_{j}+Y_{d}^{ij}Q_{i}H_{d}d^{c}_{j}+Y_{e}^{ij}L_{i}H_{d}e^{c}_{j}   \nonumber\\
&+&Y_{\nu}^{ij}L_{i}H_{u}N^{c}_{i} + Y^{ij}_{N}N^{c}_{i}N^{c}_{j}\eta_{1}+\mu^{\prime}\eta_{1}\eta_{2},
\label{superpotential}
\end{eqnarray}
where the first line represents the MSSM Superpotential using the standard notation for (s)particles while the second line includes the terms associated with the  right-handed neutrinos, $N_{i}^{c}$'s, plus the singlet Higgs fields $\eta_{1}$ and $\eta_{2}$. The $B-L$ symmetry requires $\eta_{1}$ and $\eta_{2}$ to carry $-2$ and $+2$ charges under 
$B-L$  transformations, respectively. The presence of the $N_{i}^{c}$ terms makes it possible to have Yukawa interaction terms for the neutrinos, denoted by $Y_{\nu}$. Finally, $\mu'$ stands for the bilinear mixing term between the singlet Higgs fields. 

In addition to the right-handed neutrinos and the singlet Higgs fields, the BLSSM also introduces a gauge field ($B'$) and its gaugino ($\tilde{B}'$) associated with the gauged $B-L$ symmetry, so that the appropriate Soft SUSY-Breaking (SSB) Lagrangian can be written as 
\begin{eqnarray}
-\mathcal{L}_{\rm SSB}^{{\rm BLSSM}}&=& -\mathcal{L}^{{\rm MSSM}}_{\rm SSB} +m^{2}_{\tilde{N}^{c}}|\tilde{N}^{c}|^{2} + m_{\eta_{1}}^{2}|\eta_{1}|^{2}+ m_{\eta_{2}}^{2}|\eta_{2}|^{2} +A_{\nu}\tilde{L}H_{u}\tilde{N}^{c} +A_{N}\tilde{N}^{c}\tilde{N}^{c}\eta_{1}\nonumber\\
&+& \frac{1}{2}M_{B^{\prime}}\tilde{B}^{\prime}\tilde{B}^{\prime} + M_{BB'} \tilde{B}\tilde{B}^{\prime} +B(\mu^{\prime}\eta_{1}\eta_{2}+ {\rm h.c.}).
\label{SSBLag}
\end{eqnarray}
Note that, in contrast to its non-SUSY version, the BLSSM does not allow mixing between the doublet and singlet Higgs fields through the Superpotential and SSB Lagrangian. Therefore, the scalar potential for these can be written separately and their mass matrices can be diagonalised independently. The scalar potential for the singlet Higgs fields can be derived as
\begin{equation}
V(\eta_{1} , \eta_{2})=\mu^{\prime 2}_{1}|\eta_{1}|^{2}+\mu^{\prime 2}_{2}|\eta_{2}|^{2}-\mu^{\prime}_{3}(\eta_{1} \eta_{2} + {\rm h.c.}) +\frac{1}{2}g_{BL}^{2}(|\eta_{1}|^{2}-|\eta_{2}|^{2})^{2}
\label{singpotential}
\end{equation}
and the minimisation of this potential yields Eq.~(\ref{BLmin}). Despite the non-mixing Superpotential and SSB Lagrangian, one can implement mixing between the doublet and singlet Higgs fields via $-\chi B_{\mu\nu}^{B-L}B^{Y,\mu\nu}$, where $B^a_{\mu\nu}$ is the field strength tensor of a $U(1)$ gauge field, with $a = Y,~B-L$, the hypercharge and $B-L$ charge, respectively. In the case of gauge kinetic mixing, the  covariant derivative takes a non-canonical form \cite{O'Leary:2011yq} which couples the singlet Higgs fields to the doublet ones at tree-level. Even though $\tilde{g}$ is set to zero at the GUT scale, it can be generated at the low scale through the RGEs \cite{Holdom:1985ag}.  In this basis, one finds
\be
M_Z^2\,\simeq\,\frac{1}{4} (g_1^2 +g_2^2) v^2,  ~~ ~~~~~~~  M_{Z'}^2\, \simeq\, g_{BL}^2 v'^2 + \frac{1}{4} \tilde{g}^2 v^2 ,
\ee
where $v=\sqrt{v^2_u+v^2_d}\simeq 246$ GeV and $v'=\sqrt{v'^2_1+v'^2_{2}}$ with the 
Vacuum Expectation Values (VEVs) of the
Higgs fields given by $\langle{\rm Re} H_{u,d}^0\rangle=v_{u,d}/\sqrt{2}$ and $\langle{\rm Re} ~\eta_{1,2}\rangle=v'_{1,2})/\sqrt{2}$. 
It is worth mentioning that  the mixing angle between $Z$ and $Z'$ is given by 
\be 
\tan 2 \theta'\, \simeq\, \frac{2 \tilde{g}\sqrt{g_1^2+g_2^2}}{\tilde{g}^2 + 16 (\frac{v'}{v})^2 g_{BL}^2 -g_2^2 -g_1^2}.
\ee
The minimisation conditions of the BLSSM scalar potential at tree-level lead to the following relations \cite{O'Leary:2011yq}: 
\bea
v_1' \left( m^2_{\eta_1} + \vert \mu' \vert^2 + \frac{1}{4} \tilde{g}g_{BL} (v_d^2 -v_u^2) +  \frac{1}{2} g^2_{BL} (v'^2_1 - v'^2_2) \right) - v'_2 B\mu' &=& 0 ,\\
v_2' \left( m^2_{\eta_2} + \vert \mu' \vert^2 + \frac{1}{4} \tilde{g}g_{BL} (v_u^2 -v_d^2) +  \frac{1}{2} g^2_{BL} (v'^2_2 - v'^2_1) \right) - v'_1 B\mu' 
&=& 0 .
\eea
From these equations, one can determine $\vert \mu'\vert^2$ and $B \mu'$ in terms of other soft SUSY breaking terms. Note that, with $\tilde{g}=0$, the expression of $\vert \mu'\vert^2$ takes the form of Eq. (\ref{BLmin}). 

It should  be noted here that the SUSY $B-L$ extension does not affect the chargino mass matrix, that is, the latter will be exactly the same as that of the MSSM. This is not the case for the neutralino mass matrix though. This will in fact be extended to be a $7 \times 7$ mass matrix.  This happens as a consequence of the existence of the additional neutral states $\tilde{B'}, \tilde{\eta}_1$ and $ \tilde{\eta}_2$. Thus, the neutralino mass matrix is given by
\bea
&&{\cal M}_7({\tilde B},~{\tilde W}^3,~{\tilde
	H}^0_1,~{\tilde H}^0_2,~{\tilde B'},~{\tilde{\eta}_1},~{\tilde{\eta}_2}) \equiv \left(\begin{array}{cc}
	{\cal M}_4 & {\cal O}\\
	{\cal O}^T &  {\cal M}_3\\
\end{array}\right),
\eea%
where ${\cal M}_4$ is the MSSM-type neutralino mass matrix and
${\cal M}_{3}$ is the additional $3\times 3$ neutralino mass matrix,
which is given by%
\be%
{\cal M}_3 = \left(\begin{array}{ccc}
	M_{B'} & -g_{_{BL}}v'_1  & g_{_{BL}}v'_2 \\
	-g_{_{BL}}v'_1 & 0 & -\mu'  \\
	g_{_{BL}}v'_2 & -\mu' & 0\\
\end{array}\right).
\label{mass-matrix.1} \ee
In addition, the off-diagonal matrix ${\cal O}$ is given by
\be%
{\cal O} = \left(\begin{array}{ccc}
	M_{BB'} &~~~0~~~& 0 \\
	0 & 0 & 0  \\
	-\frac{1}{2}\tilde{g}v_d &~~~0~~~& 0\\
	\frac{1}{2}\tilde{g}v_u&~~~0~~~&0\\
\end{array}\right).
\label{mass-matrix.2} \ee
(Note that the off-diagonal matrix elements vanish identically if $\tilde{g}=0$ and $M_{BB'} = 0$). One can then diagonalise the real matrix ${\cal M}_{7}$ with
a symmetric mixing matrix $V$ such that
\be V{\cal
	M}_7V^{T}={\rm diag}(m_{\tilde\chi^0_k}),~~k=1,\dots,7.\label{general} \ee In
these conditions, the LSP has the following decomposition 
\be \label{cm_neuComp}
\tilde\chi^0_1=V_{11}{\tilde B}+V_{12}{\tilde
	W}^3+V_{13}{\tilde H}^0_d+V_{14}{\tilde
	H}^0_u+V_{15}{\tilde B'}+V_{16}{\tilde{\eta}_1}+V_{17}{\tilde{\eta_2}}. 
\ee 

If the LSP  is then  considered as a candidate for DM, each species in the above equation, if dominant, 
leads to its own phenomenology that can possibly be distinguished in direct detection experiments. For example, to achieve 
the  correct relic density of Bino-like DM is challenging, since its abundance is usually so high over the fundamental parameter space that one needs to identify several annihilation and/or coannihilation channels to reduce its density down to the
Wilkinson Microwave Anisotropy Probe
 (WMAP) \cite{Hinshaw:2012aka} or Planck \cite{Ade:2015xua} measurements. Since this DM state interacts through the hypercharge, its scattering with  nuclei has a very low cross section. Conversely, the largest cross section in DM scattering can be obtained when DM is  Higgsino-like, since it interacts with the quarks through the Yukawa interactions. Since the BLSSM sector offers significant interference in the neutralino sector, this may also drastically change the DM kinematics. In contrast to a Bino, the $\tilde{B'}-$ino interacts more strongly depending on the $B-L$ gauge coupling. Despite the severe mass bound on the $Z'$, there is no specific bound on $m_{\tilde{B}'}$, so that it can be even as low as 100 GeV \cite{Khalil:2015wua}. In this context, one can expect the LSP neutralino to be mostly formed by $\tilde{B}'$ and its cross section in its scattering with nuclei can be very large, in contrast to the Bino case. In addition to $\tilde{B}'$, the LSP neutralino can be formed by the singlet Higgsinos (also dubbed Bileptinos due to their $L=\pm2$ lepton charge). In this case, it is challenging for their abundance
to be compatible with the experimental results. The reduction through the coannihilation channels involving SUSY particles arises from the gauge kinetic mixing, which is restricted to be moderate. If its mass is nearly degenerate with that of the $\tilde{B}'$ state, they can significantly coannihilate. Also, a singlet Higgsino yields low cross section in DM scattering experiments.
Besides the neutralinos, one can also consider the sneutrino as a DM candidate when it is the LSP, of course.
In this case, the extended sector of the BLSSM involves twelve states coming from the Superpartners of the left- and the right-handed neutrinos. In a Charge and Parity (CP)-conserving framework the states entering the sneutrino mixing matrix can be expressed by separating their scalar and pseudo-scalar components
\bea \label{SneutrinoCP}
\tilde{\nu}_i = \frac{\sigma_L{}_i + i \phi_L{}_i}{\sqrt{2}} \,,\quad
\tilde{N}_i = \frac{\sigma_R{}_i + i \phi_R{}_i}{\sqrt{2}}.
\eea 
The breaking of  $B-L$ generates an effective mass term through $Y^{ij}_{N}N^{c}_{i}N^{c}_{j}\eta_{1}$ causing a mass splitting between the CP-even and CP-odd
sector. Therefore, in terms of Eq. (\ref{SneutrinoCP}), the corresponding $12\times12$ mass matrix is reduced to two different $6\times6$ blocks
\bea
{\cal M}^{2\,\sigma}(\sigma_L, \sigma_R)  \equiv \left(\begin{array}{cc}
	{\cal M}^{2\,\sigma}_{LL} & {\cal M}^{2\,\sigma}_{LR}\\
	{\cal M}^{2\,\sigma}_{LR}{}^T &  {\cal M}^{2\,\sigma}_{RR}\\
\end{array}\right)\,,\quad
{\cal M}^{2\,\phi}(\phi_L, \phi_R) \equiv \left(\begin{array}{cc}
	{\cal M}^{2\,\phi}_{LL} & {\cal M}^{2\,\phi}_{LR}\\
	{\cal M}^{2\,\phi}_{LR}{}^T &  {\cal M}^{2\,\phi}_{RR}\\
\end{array}\right).
\eea
Such differences between CP-even and CP-odd sectors do not involve the left components with ${\cal M}^{\sigma}_{LL}$ and ${\cal M}^{\phi}_{LL}$
described  by the common form ${\cal M}^{2}_{LL}$
\bea
{\cal M}^{2}_{LL}{}^{i,j} \equiv \frac{\delta^{i,j}}{8} \left( \left( g_1^2 + g_2^2 + \tilde{g}\left(g_{BL} + \tilde{g}\right) \right) \delta_{H} + \left(g_{BL} + \tilde{g}\right)\delta_{\eta} \right)
+ \frac{1}{2} v_u^2 \left( Y_{\nu}^T Y_{\nu}\right)^{i,j} +  m_l^2{}^{i,j}, \nn \\
\eea
where we have introduced  $\delta_{\eta} = v'^2_{1} - v'^2_{2}$ and $\delta_{H} = v^2_d - v^2_u$ .
For the submatrices  ${\cal M}^{2\,\sigma}_{RR}$ and ${\cal M}^{2\,\phi}_{RR}$ we have instead
\bea
{\cal M}^2_{RR}{}^{i,j} &\equiv& - \frac{\delta^{i,j}}{8}\,g_{BL} \left(\tilde{g} \delta_H + 2 g_{BL} \delta_{\eta} \right) + \frac{1}{2}\,v_u^2 \left(Y_{\nu} Y_{\nu}^T\right)^{i,j}
+  m_{\tilde{N}}^2{}^{i,j} + 2\,v'^2_{1}\, \left(Y_N^2\right)^{i,j} \nn \\  &\mp&  \sqrt{2} \left( v'_{2}\, \mu' Y_{N}^{i,j} - v'_{1}\, A_{N}^{i,j} \right)  
\eea
while the left-right sneutrino mixing is ruled by the matrices
\bea
{\cal M}^2_{LR}{}^{i,j} &\equiv& \frac{1}{2}\left( - \sqrt{2}\,v_{d} \mu Y_{\nu}^{i,j} + v_u\,\sqrt{2}\, A_{\nu}^{i,j} \pm 2 v_u\,v'_{1}\, \left(Y_{N}Y_{\nu} \right)^{i,j} \right),
\eea
with upper(lower) signs corresponding to CP-even(odd) cases. The parameter $Y_{\nu}$ and the corresponding trilinear term $A_{\nu}$ determine the mixing between
the left and  right components. In our setup, $Y_{\nu}$ is negligible and can safely be set to zero already at the GUT scale, as it is the case also for the boundary condition of $A_{\nu}$.
The resulting  $12\times12$ sneutrino mass matrix is consequently unable to mix the left- and right-handed components as the CP-even and CP-odd parts of  a sneutrino state
will be completely determined by assigning its CP value and the chirality of its Supersymmetric partner.

\section{Renormalisation Group Equations}
\label{sec:rge}

The presence of an extra Abelian gauge group introduces a distinctive feature, the gauge kinetic mixing, through a renormalisable and gauge invariant operator $\chi B^{\mu\nu}B'_{\mu\nu}$ of the two Abelian field strengths. 
Moreover, off-diagonal soft breaking terms for the Abelian gaugino masses are also allowed.
This effect is completely novel with respect to the MSSM or other Supersymmetric models in which only a single $U(1)$ factor is considered. 
Even if the kinetic mixing is required to vanish at a given scale, the RGE evolution inevitably reintroduces it alongside the running, unless a particular field content and charge assignment are enforced. If the two Abelian gauge factors emerge from the breaking of a simple gauge group, the kinetic mixing is absent at that scale. For this reason, arguing that the BLSSM could be embedded into a wider GUT scenario (the matter content of the BLSSM, which includes three generations of right-handed neutrinos, nicely fits into the 16-D spinorial representation of $SO(10)$), we require the vanishing of the kinetic mixing at the GUT scale. As we stated above, we nevertheless end up with a non-zero kinetic mixing at low scales affecting the $Z'$ interactions as well as the Higgs and the neutralino sectors \cite{O'Leary:2011yq}. \\
Instead of working with a non-canonical kinetic Lagrangian in which the kinetic mixing $\chi$ appears, it is more practical to introduce a non-diagonal gauge covariant derivative with a diagonal kinetic Lagrangian. The two approaches are related by a gauge field redefinition and are completely equivalent. In this basis the covariant derivative of the Abelian fields takes the form $\mathcal D_\mu = \partial_\mu - i Q^T G A_\mu$, where $Q$ is the vector of the Abelian charges, $A$ is the vector of the Abelian gauge fields and $G$ is the Abelian gauge coupling matrix with non-zero off-diagonal elements. The matrix $G$ can be recast into a triangular form with an orthogonal transformation $G \rightarrow G O^T$ \cite{Coriano:2015sea}. With this parametrisation, the three independent parameters of $G$ are explicitly manifest and correspond to the Abelian couplings, $g_1$, $g_{BL}$ and $\tilde g$, describing, respectively, the hypercharge interactions, the extra $B-L$ ones  and the gauge kinetic mixing. Differently from  the MSSM case, the Abelian gaugino mass term is replaced by a symmetric matrix with a non-zero mixed mass term $M_{BB'}$ between the $B$ and $B'$ gauginos. Coherently with our high energy unified embedding, we choose $M_{BB'} = 0$ at the GUT scale. Notice that the Abelian gaugino mass matrix $M$ is affected by the same rotation $O$ and in the basis in which $G$ is triangular and $M$ transforms through $M \rightarrow O M O^T$. \\
We have performed a RGE study of the BLSSM assuming gauge coupling unification and mSUGRA boundary conditions at the GUT scale.
The two-loop RGEs have been computed with SARAH \cite{Staub:2013tta} and fed into SPheno \cite{Porod:2003um} which has been used for the spectrum computation and for the numerical analysis of the model. Here we show the one-loop $\beta$ functions of the gauge couplings highlighting the appearance of the kinetic mixing contributions
\bea
\label{eq:RGEgauge}
\beta^{(1)}_{g_1} &=& \frac{33}{5} g_1^3, \nn \\
\beta^{(1)}_{g_{BL}} &=& \frac{3}{5} g_{BL} \left( 15 g_{BL}^2  + 4 \sqrt{10} g_{BL} \, \tilde g + 11 \tilde g^2  \right), \nn \\
\beta^{(1)}_{\tilde g} &=& \frac{3}{5} \tilde g \left( 15 g_{BL}^2  + 4 \sqrt{10} g_{BL} \, \tilde g + 11 \tilde g^2 \right) + \frac{12 \sqrt{10}}{5} g_1^2 g_{BL}, \nn \\
\beta^{(1)}_{g_2} &=& g_2^3, \nn \\
\beta^{(1)}_{g_3} &=& -3 g_3^3, 
\eea
where we have adopted the GUT normalisations $\sqrt{3/5}$ and $\sqrt{3/2}$, respectively, for the $U(1)_Y$ and  $U(1)_{B-L}$ gauge groups.
At one-loop level the expressions of the $\beta$ functions of $g_1$, $g_2$ and $g_3$ are the same as those of the MSSM with differences appearing at two-loop order only. 
Notice that the term responsible for the reintroduction of a non-vanishing mixing coupling $\tilde g$ along the RGE running, even if absent at some given scale, is the last term in $\beta^{(1)}_{\tilde g}$. We recall again that the kinetic mixing is a peculiar feature of Abelian extensions of the SM and their Supersymmetric versions, admissible only between two or more $U(1)$ gauge groups. 

Assuming gauge coupling unification at the GUT scale, the RGE analysis provides the results $\tilde g \simeq -0.144$ and $g_{BL} \simeq 0.55$ with $M_{\textrm{GUT}} \simeq 10^{16}$ GeV, which are controlled by the leading one-loop $\beta$ functions given in Eq.~(\ref{eq:RGEgauge}). The spread of points around these central values, less than 1\% for $g_{BL}$ and 5\% for $\tilde g$, is only due to higher-order corrections, namely two-loop running and threshold corrections. 

\begin{figure}[!t]
\centering
\includegraphics[scale=0.5]{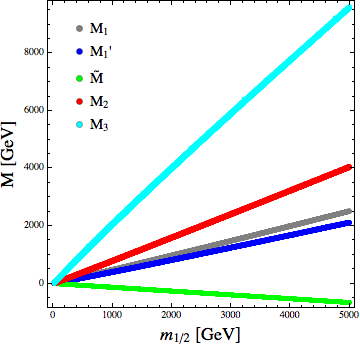} 
\caption{Gaugino masses at the SUSY  scale as a function of the GUT $m_{1/2}$ mass. Here,  both gauge coupling and soft mass unification have been assumed. \label{fig:gauginomass}}
\end{figure}
The running of the gaugino masses is directly linked to that of the gauge couplings. In the Abelian sector and at one-loop, the Abelian gaugino mass matrix $M$ evolves with
\bea
\beta_M = M G^T Q^2 G + G^T Q^2 G M = M G^{-1} \beta_G + G^{-1} \beta_G M, 
\eea
where $Q = \sum_p Q_p Q_p^T$, with $Q_p$ the vector of the Abelian charges of the $p$ particle.
Exploiting the structure of the $\beta$ functions of the gaugino masses, a simple relation is obtained, $M_i/m_{1/2} = g_i^2/g_{\rm GUT}^2$, for non-Abelian masses at one-loop order. In the Abelian sector, due to the presence of the mixing, the previous equation is replaced by a matrix relation. Indeed, from the product $G M^{-1} G^T$, which remains constant along the RGE evolution, one finds the Abelian gaugino mass matrix $M/m_{1/2} = G^T G / g_{\rm GUT}^2$. We show in Fig.~\ref{fig:gauginomass} the  dependence of the gaugino masses as a function of the GUT gaugino mass $m_{1/2}$. The hierarchy is obviously controlled by the size of the gauge couplings at low scale.

The one-loop $\beta$ functions of the soft masses of the scalar fields $H_u, H_d$ and $\eta_1, \eta_2$ are given by
\bea
\beta_{m_{H_u}^2} &=& - \frac{6}{5} \left( g_1^2 (M_1^{2} + \tilde M^2 ) + \tilde g^2  ({M'_1}^{2}  + \tilde M^2 ) + 2 g_1 \tilde g ( M_1 + M'_1) \tilde M \right) - 6 g_2^2 M_{W}^2, \nn \\
&-& 3 (g_1^2 + \tilde g^2) \sigma_1  - \frac{3 \sqrt{10}}{4} g_{BL} \tilde g \sigma_2 +6 \left( m_{H_u}^2  + m_{q_{33}}^2  + m_{u_{33}}^2  \right) Y_t^2 + 6 T_t^2 \\
\beta_{m_{H_d}^2} &=& - \frac{6}{5} \left( g_1^2 (M_1^{2} + \tilde M^2 ) + \tilde g^2  ({M'_1}^{2}  + \tilde M^2 ) + 2 g_1 \tilde g ( M_1 + M'_1) \tilde M \right) - 6 g_2^2 M_{W}^2,  \nn \\
&+& 3 (g_1^2 + \tilde g^2) \sigma_1  + \frac{3 \sqrt{10}}{4} g_{BL} \tilde g \sigma_2 +6 \left( m_{H_d}^2  + m_{q_{33}}^2  + m_{d_{33}}^2  \right) Y_b^2 + 6 T_b^2 \\
\beta_{m_{\eta_1}^2} &=& -12 {g_{BL}}^{2} ({M'_1}^{2} + \tilde M^2 ) + 4 m_{\eta_1}^2 \textrm{tr}(Y_N^2) + 4 \textrm{tr}(T_{Y_N}^2) + 8  \textrm{tr}(m_{\nu_R}^2 Y_N^2), \nn  \\
&+& 3 \sqrt{\frac{2}{5}} g_{BL} \tilde g \sigma_1 + \frac{3}{2} {g_{BL}}^2 \sigma_2, \\
\beta_{m_{\eta_2}^2} &=& -12 {g_{BL}}^{2} ({M'_1}^{2} + \tilde M^2 ) - 3 \sqrt{\frac{2}{5}} g_{BL} \tilde \, g \sigma_1 - \frac{3}{2} {g_{BL}}^2 \sigma_2,
\eea
where, for the sake of simplicity, we have neglected all the Yukawa couplings but top- and bottom-quark $Y_t, Y_b$ and the heavy-neutrinos $Y_N$. We have also assumed real parameters. The gaugino masses $M_1, M'_1$ and $\tilde M$ are obtained from $M_B, M_{B'}$ and $M_{BB'}$ through the transformation $OMO^T$.
The coefficients $\sigma_{1,2}$ are defined as
\bea
\sigma_1 &=& m_{H_d}^2 - m_{H_u}^2 - \textrm{tr}( m_d^2) - \textrm{tr}( m_e^2) + \textrm{tr}( m_l^2) - \textrm{tr}( m_q^2) + 2  \textrm{tr}( m_u^2), \nn \\
\sigma_2 &=& 2 m_{\eta_1}^2 - 2 m_{\eta_2}^2 + \textrm{tr}( m_d^2) - \textrm{tr}( m_e^2) + 2 \textrm{tr}( m_l^2) - 2 \textrm{tr}( m_q^2) +  \textrm{tr}( m_u^2) - \textrm{tr}( m_{\nu_R}^2)
\eea
and are found to be RGE invariant combinations of the soft SUSY masses. Assuming  unification conditions at the GUT scale,  $\sigma_{1,2}$ remain zero along all the RGE evolution. 
Being $\beta_{m_{\eta_2}^2}$ characterised only by negative contributions proportional to the Abelian gaugino masses, the corresponding soft mass $m_{\eta_2}^2$ will increase and remain positive during  the run from the GUT to the EW scale. The same feature is shared by $m_{H_d}^2$ except for some particular values of the gaugino and soft scalar masses at the GUT scale for which the $Y_b$ Yukawa coupling contribution (of the $b$-quark) to $\beta_{m_{\eta_2}^2}$ is not negligible. The spontaneous symmetry breaking of EW and B-L, requiring negative $m^2_{H_u}$ and $m^2_{\eta_1}$, can be realised radiatively, which is a nice feature in both MSSM and BLSSM. Namely, even though there is no spontaneous symmetry breaking at a high scale, the large top-quark Yukawa coupling $Y_t$ and its trilinear soft term $A_t$ can drive $m^2_{H_u}$ negative through its RGE evolution, which triggers spontaneous EWSB. Similarly, a sufficiently large neutrino Yukawa coupling $Y_N$ and corresponding trilinear soft term $A_n$ turn $m^2_{\eta_1}$ negative in its RGE evolution and break the $B-L$ symmetry spontaneously.

In general only one of the three components of the diagonal $Y_N$ matrix is required to be large in order to realise the spontaneous symmetry breaking of the extra Abelian symmetry, thus providing a heavy and two possible lighter heavy-neutrino states. Notice also that the elements of the low scale values of the $Y_N$ matrix cannot be taken arbitrary large otherwise a Landau pole is hit before the GUT scale. A close inspection of the one-loop $\beta$ function of the heavy-neutrino Yukawa coupling
\bea
\beta_{Y_N} = 8 Y_N Y_N^* Y_N + 2 \textrm{tr} (Y_N Y_N^*) Y_N - \frac{9}{2} {g_{BL}}^2 Y_N,
\eea
where we have neglected the negligible contribution of the light-neutrino Yukawa coupling $Y_\nu$, shows that $Y_N \gtrsim 0.5$ spoils indeed the perturbativity of the model at the GUT scale or below.

\section{Collider and Dark Matter Constraints}
\label{sec:colliderdm}
To investigate the viability of the BLSSM parameter space, with mSUGRA boundary conditions, we have challenged its potential signatures against two sets of experimental constraints. 
To the first set belong different bounds coming from collider probes which have been used in building the scan procedure. These form a varied  set of requirements
affecting our choice of the $Z'$ benchmark mass as well as the character of the acceptable low-scale  particle spectrum. 
As already stated, stringent constraints come from LEP2 data via EW Precision Observables (EWPOs) and from 
 Run 2 of the LHC through a signal-to-background analysis using Poisson statistics to extract a 95\% Confidence Level (CL) bound in the di-lepton channel. The CL has been extracted at the LHC with $\sqrt{s} = 13$ TeV and $\mathcal L = 13.3$ fb$^{-1}$, updating the analysis presented in \cite{Accomando:2016sge}. We have taken into account the $Z'$ signal and its interference with the SM background and included efficiency and acceptance for both the electron and muon channels as described in \cite{Khachatryan:2016zqb}.
Such studies affect the extended gauge sector $\left(\tilde{g}, g_{BL}, M_{Z'}\right)$ in a way that, in all safety, allow us to select the value
$M_{Z'} = 4$ TeV for all  magnitudes of gauge couplings and $Z'$ total width (in the range 30--45 GeV) met in the RGE evolution. Notice that the BLSSM supplied with unification conditions at the GUT scale provides a very narrow $Z'$ width with a $\Gamma_{Z'}/M_{Z'}$ ratio reaching $~1\%$ at most. Thus, this is unlike the results of \cite{Abdallah:2015hma,Abdallah:2015uba}, which were indeed obtained without any universality conditions. 
Such a $Z'$ mass value completes the independent parameters that feed our scan and which in turn provides a BLSSM low-energy spectrum. 
It is at this stage that we can force the exclusion bounds coming from LEP, Tevatron and LHC linked to the negative searches of scalar degrees of freedom and to the correct reproduction of the measured Higgs signal strength around $125$ GeV.
More precisely, from our scan it is possible to extract the masses and the Branching Ratios (BRs) of all the (neutral and charged) scalars plus their effective couplings to SM fermions and bosons. This information is then processed into {\texttt HiggsBounds} \cite{Bechtle:2008jh,Bechtle:2011sb,Bechtle:2013wla,Bechtle:2015pma} which, considering all 
the available collider searches, expresses whether a parameter point has been excluded at $95\%$ CL or not.

This analysis establishes a first solid sieve by reducing a considerable number of acceptable points, among those with successful EW and $U(1)_{B-L}$ symmetry breaking, as 
obtained from the GUT parameters scan. 
Over such points, the compatibility fit of the generated Higgs signal strengths with the ones measured at LHC is taken into account by {\texttt HiggsSignals} 
\cite{Bechtle:2013xfa}, which provides the corresponding $\chi^2$. By asking for a $2\sigma$ interval around the minimum $\chi^2$ generated, we obtain a further constraint over the 
parameter space investigated. 

The second set of bounds that we considered emerges from the probe of DM signatures which are a common and natural product of many SUSY models. 
Among these, the BLSSM stands out for both theoretical and phenomenological reasons that make the study of its DM aspects particularly worthwhile. 
The presence of a gauged $B-L$ symmetry, being broken by the scalar fields $\eta_1$ and $\eta_2$, as they are charged under $B-L$ \cite{FileviezPerez:2010ek},  provides a local origin 
to the discrete $R-$symmetry that is usually imposed ad-hoc to prevent fast proton decay. Consequently, the BLSSM embeds the stability of the LSP through its gauge structure, as it does for the produced DM density. 

From the phenomenological side, the BLSSM, like  the MSSM, has the neutralino as a possible cold DM candidate. The presence of additional neutral degrees of freedom drastically changes its properties with  respect to the corresponding MSSM ones, which is mostly Bino in GUT constrained models, possibly 
giving the necessary degrees of freedom to accommodate  the measured DM evidences. Moreover, the BLSSM also envisages a scalar LSP in its spectrum, generated by the superpartners of the 
six Majorana neutrinos, which may also be the origin of a cold DM relic. 

For every possible low energy spectrum obtained, the LSP provided by the BLSSM will participate in the early thermodynamical evolution of the universe.
After an initial regime of thermal equilibrium with the SM particles, decoupling takes place once the DM annihilation rate becomes slower than the Universe 
expansion. 
This process would result in the relic density lasting until now. 
Consequently, a crucial test of the cosmological viability of the BLSSM  is enforced by requiring the relic abundance generated not to overclose the Universe
by exceeding the measured current value of the DM relic density 
\begin{equation} \label{PLANCK}
\Omega h^2 = 0.1187 \pm 0.0017({\rm stat}) \pm 0.0120({\rm syst}) 
\end{equation}
as measured by the Planck Collaboration \cite{Ade:2015xua}.

The requirement to reproduce the measured relic density would finally highlight the region of the parameter space where
the model is able to solve the DM puzzle.
The computation of the DM abundance is achieved by solving the evolution numerically with 
{\texttt MicrOMEGAs} \cite{Belanger:2006is,Belanger:2013oya}, which collects the amplitudes for all the annihilation, as well as coannihilation, processes. 
Another source of constraints, which cannot be neglected due to the recent increase in precision reached by the LUX collaboration  \cite{Akerib:2016lao,Akerib:2016vxi},
is linked to the 
direct searches intended to detect DM signatures coming from DM scatterings with nuclei. 
We have tested the BLSSM spectrum against the challenging upper limit on the Spin Independent (SI) component of the  LSP-nucleus scattering. 
The zeptobarn order of magnitude, reached in the recent upgrade of the DM-nucleus cross section bound, will have an interesting interplay with the parameter 
space analysed to test the surviving ability of the BLSSM against stringent exclusions. 

The DM scenarios provided represent a peculiar signature of the model, with characteristic degrees of freedom playing a key role
in drawing a rich DM texture. 
As already stated, the BLSSM has two candidates for cold DM as it is possible to have, other than the neutralino, also a heavy stable sneutrino.
The extended neutral sector, consequence of the inclusion of an extra $B-L$ gauge factor, enlarges the neutralino components with three new states (two coming from Bileptinos and one from BLino) as seen in Eq. (\ref{cm_neuComp}).
To  study the behaviour of the neutralinos we may consider the following classification 
\begin{center}
\begin{tabular}{ll}
$V_{11}^2 > 0.5$ & Bino-like,\\
$V_{12}^2 > 0.5$ & Wino-like,\\
$V_{13}^2 + V_{14}^2 > 0.5 $& Higgsino-like,\\
$V_{15}^2 > 0.5 $& BLino-like,\\
$V_{16}^2 + V_{17}^2 > 0.5$ & Bileptino-like,\\
Neither of the previous cases & Mixed. 
\end{tabular}
\end{center}
In this scheme the nature of the neutralino is identified with the interaction eigenstate that makes up for more than half of its content. 

\begin{figure}[ht!]
	\centering
	\includegraphics[scale=0.4]{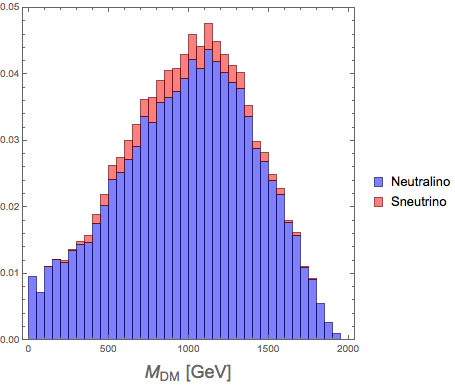}
	\includegraphics[scale=0.4]{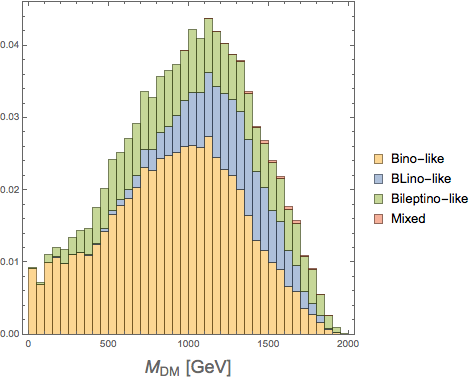}
	\caption{(a) The normalised distribution of the neutralino and sneutrino types found in our scan.
	         (b) The normalised distribution of the different types of LSP found in our scan. The histograms are  stacked.}{}
	\label{composition}
\end{figure}
For all the points generated in our scan, in agreement with the constraints from Higgs searches,
the LSP will, in the majority of cases, results in a fermionic DM candidate with mass below 2 TeV, see Fig.~\ref{composition}(a).
The sneutrino will instead be a subdominant option over our entire set of points.
It is interesting to explore the composition of the sneutrino LSP written  in terms of CP eigenstates and left-right parts. 
This is relevant to appreciate the chances to survive the direct detection probes of DM, with a left-handed sneutrino having a dangerously enhanced 
scattering rate against nuclei \cite{Falk:1994es} due to $Z$ mediation.
\begin{figure}[ht!]
	\centering
	\includegraphics[scale=0.4]{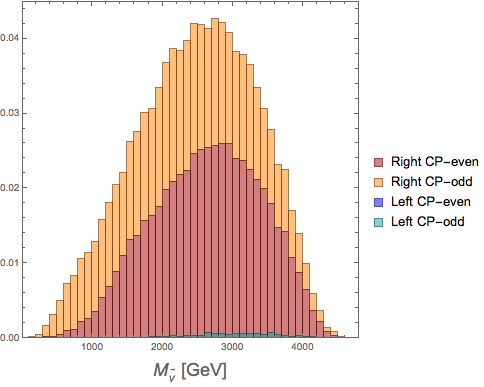}
	\caption{Composition of the lightest sneutrino for the set of points in agreement with the constraints from {\texttt HiggsBounds} and {\texttt HiggsSignals}. 
	Histogram is of stacked type with normalised heights.}
	\label{fig:SneuHist}
\end{figure}
Fig.~\ref{fig:SneuHist} indicates that the lightest sneutrino can sizeably be left-handed only above $\sim$ 2 TeV. The complementary region is the only one
where the sneutrino can compete against the neutralino as a possible LSP clarifying why the LSP sneutrino met in our constrained BLSSM will always be a \emph{right}-handed sneutrino. 
Following the previous classification, a Bino-like neutralino will be more common to encounter as the BLSSM favourite LSP, but, as typical 
features of the model, also states of BLino and Bileptino nature are often met, see Fig.~\ref{composition}(b).
Notably, no Higgsino-like neutralino are found while the Wino possibility is a most rare one, which requires very tuned conditions over the parameter space to be
produced in a sizeable amount. Given our uniform treatment over the boundary conditions, we will not consider this case though.

\section{Fine-Tuning Measures}
\label{sec:fine_tuning}
We introduce measures of FT in this section to compare BLSSM and MSSM in respect of naturalness. FT is not a physical observable, but it is rather an indication for an unknown mechanism, which is missing in the model under concern. Its quantitative values, then, can be interpreted as the effectiveness of the missing mechanisms over the low scale results. In this context, the model may cover most of the whole BSM physics, when FT is small.

There are many alternatives for a quantitative measure of FT \cite{Anderson:1994dz,Anderson:1994tr,Anderson:1995cp,Anderson:1996ew,Ciafaloni:1996zh,Chan:1997bi,Barbieri:1998uv,Giusti:1998gz,Casas:2003jx,Casas:2004uu,Casas:2004gh,Casas:2006bd,Kitano:2005wc,Athron:2007ry,Baer:2012up}, which are commonly based on the change in the $Z$-boson mass. Its measure (denoted by $\Delta$) equals the largest of these changes defined as \cite{Ellis:1985yc,Barbieri:1987fn}
\begin{equation}
\Delta={\rm Max} \left| \frac{\partial \ln v^2}{\partial \ln a_i}\right| =  {\rm Max} \left| \frac{a_i}{v^2} \frac{\partial v^2}{\partial a_i } \right|  = {\rm Max} \left| \frac{a_i}{M_Z ^2} \frac{\partial M_Z ^2}{\partial a_i} \right|.
\label{eq:BGFT}
\end{equation}
When viewing a parameter space, a particular point has a low FT if the $Z$ mass does not largely change when deviating from its position. A natural model will, therefore, possess large regions of viable parameter space with low FT values. Having this feature in a particular model will make it more attractive a prospect. Our goal here is to find allowed regions of parameter space for the BLSSM with a similar (or better) level of FT to the MSSM, so the models may be of comparable naturalness.
In this paper, we apply this same measure in two different scenarios (high- and low-scale parameters) for both the MSSM and BLSSM. We will proceed by explaining the procedure for the two models. Firstly, we minimise the Supersymmetric Higgs potential (after EWSB) with respect to the EW VEVs. We do this at loop level. These minimisation conditions are called tadpole equations and may be solved to find a relation for the $Z$-mass and SUSY-scale  parameters. At this point, we have two choices: to use these SUSY-scale  parameters or to relate these to high-scale (GUT) ones and use those. We use a fundamentally different treatment of the loop contributions to the FT in both cases. For the GUT-FT, we treat loop corrections as dependent on the EW VEV, as done in \cite{Ross:2017kjc}, which will eventually reduce the FT  value by up to a factor of $\sim$ 2. For the SUSY-scale  parameters, we treat these as independent of the EW VEV, as done in \cite{Baer:2012up}, which will not affect the FT between tree and loop-level.
We begin first by discussing the high and low scale scenarios for the MSSM, and proceed to extend this discussion to the BLSSM. 

For the GUT-FT in the MSSM, our high-scale parameters are: the unification masses for scalars ($m_0$)  and gauginos ($m_{1/2}$), the universal trilinear coupling ($A_0$), the $\mu$ parameter and the quadratic soft SUSY term ($B\mu$),
\be
a_i = \left\lbrace m_0 , ~ m_{1/2},~ A_0,~ \mu ,~ B \mu  \right\rbrace.
\ee
The GUT-FT will compare the naturalness at high scale, but two models with similar measures here may have large differences at the SUSY-scale. To test whether the BLSSM and MSSM have a similar FT at both GUT and SUSY-scale, we will consider a low-scale FT. To do this, we begin with the relation for the $Z$-mass and SUSY-scale  parameters,
\be
\frac{1}{2} M_Z^2= \frac{(m_{H_d}^2 + \Sigma_d ) - (m_{H_u}^2 + \Sigma_u) \tan^2 \beta}{\tan^2 \beta -1} -  \mu^2 ,
\label{eq:min_pot_MSSM} 
\ee
where
\be
\Sigma_{u,d} = \frac{\partial \Delta V}{\partial v_{u,d} ^2}.
\ee
Unlike in the GUT-FT case, we treat the loop corrections as independent of the EW VEV, as in \cite{Baer:2012up}. If we substitute this expression into Eq.~(\ref{eq:BGFT}) and use the low-scale parameters $a_i = \large{\lbrace}m_{H_d} ^2$, $m_{H_u} ^2$, $\mu ^2$, $\Sigma_u$, $\Sigma _d \large{\rbrace}$, one will find \cite{Baer:2012up}
\begin{equation}
\Delta_{\rm SUSY}\equiv {\rm Max}(C_{i})/(M_{Z}^{2}/2)~,
\label{FT}
\end{equation}
where
\begin{equation}
\hspace{0.3cm} C_{i}=\left\lbrace \begin{array}{lllll} C_{H_{u}} &=  \left| m_{H_{u}}^{2} \dfrac{\tan ^2 {\beta}}{(\tan ^2 {\beta} -1)} \right|~, ~~~~~ C_{H_{d}} =  \left| m_{H_{d}}^{2} \dfrac{1}{(\tan ^2 {\beta} -1)} \right|,
\\ & & &\\
C_{\mu} &=  \left| \mu^{2} \right|, ~~~~~ C_{\Sigma_u}= \left| \Sigma_u \dfrac{\tan ^2 {\beta}}{(\tan ^2 {\beta} -1)} \right|, ~~~~~C_{\Sigma_{d}} =  \left| \Sigma_d \dfrac{1}{(\tan ^2 {\beta} -1)} \right|. 
\end{array}\right.
\label{eq:FT_MSSM_C}
\end{equation}

We now turn to the BLSSM. For the GUT-FT, we follow the same universal parameters as the MSSM, but with two additional terms, relating to the $\mu '$ parameter and the corresponding quadratic soft SUSY term, $B \mu '$, so that all of our high scale terms are:
\be
a_i = \left\lbrace m_0 , ~ m_{1/2},~ A_0,~ \mu ,~ B \mu ,~  \mu',~ B \mu'  \right\rbrace.
\ee
We may also follow our previous procedure to find a SUSY-scale FT (SUSY-FT) for the BLSSM. By minimising the scalar potential, we find (at loop level),
\be
\frac{Mz^2}{2}=\frac{1}{X}\left( \frac{ m_{H_d}^2 + \Sigma _{d} }{ \left(\tan ^2(\beta
)-1\right)}-\frac{ (m_{H_u}^2 + \Sigma _u) \tan ^2(\beta )}{
\left(\tan ^2(\beta )-1\right)} + \frac{\tilde{g} M_{Z'}^2 Y}{4 g_{BL}
}- \mu ^2 \right), \label{eq:blssm_mz}
\ee
where 
\begin{equation}
X= 1 + \frac{\tilde{g}^{2}}{(g_{1}^{2}+g_{2}^{2})}+\frac{\tilde{g}^{3}Y}{2g_{BL}(g_{1}^{2}+g_{2}^{2})},
\end{equation}
and
\be
Y= \frac{\cos(2\beta ')}{\cos (2\beta)} = \frac{\left(\tan^2 {\beta} +1\right) \left(1-\tan^2 {\beta '} \right)}{\left(1-\tan ^2 {\beta } \right) \left(\tan ^2 {\beta '}
	+1\right) }
\ee
In the limit of no gauge kinetic mixing ($\tilde{g}\rightarrow 0$), this equation reproduces the MSSM minimised potential 
of Eq. (\ref{eq:min_pot_MSSM}).
Our SUSY-FT parameters for the BLSSM are thus
\begin{equation}
\hspace{0.3cm} C_{i}=\left\lbrace \begin{array}{llll} C_{H_{u}} &=  \left| \dfrac{m_{H_{u}}^{2}}{X} \dfrac{\tan ^2 {\beta}}{(\tan ^2 {\beta} -1)} \right|~,  C_{H_{d}} =  \left| \dfrac{m_{H_{d}}^{2}}{X} \dfrac{1}{(\tan ^2 {\beta} -1)} \right|,  C_{\Sigma_{d}} =  \left| \dfrac{\Sigma_d}{X} \dfrac{1}{(\tan ^2 {\beta} -1)} \right|
\\ & & &\\
 C_{\Sigma_{u}} &=  \left| \dfrac{\Sigma_u}{X} \dfrac{\tan ^2 {\beta}}{(\tan ^2 {\beta} -1)} \right|,
 ~C_{\mu} = \left|  \dfrac{\mu^{2}}{X} \right|,
 ~C_{Z'} =  \left| M_{Z'}^{2}\dfrac{\tilde{g}Y}{4 g_{BL} X} \right| . 
\end{array}\right.
\label{FTC}
\end{equation}
These equations resemble those of the MSSM SUSY-FT, but now with a factor of $1/X$. In addition, we have a contribution from the $Z'$ mass and BLSSM loop factors. Considering the heavy mass bound on $M_{Z'}$, its contribution could be expected much larger than the other terms in Eq.~(\ref{eq:blssm_mz}), which would worsen the required FT at the low scale. However, a significantly large $M_{Z'}$ severely constrains the VEVs of the singlet Higgs fields as $\tan\beta' \sim 1$ \cite{O'Leary:2011yq} and, hence, $Y$ yields a very stringent suppression in $C_{Z'}$. Note that, even though the trilinear $A$-terms are not included in determining the FT, their effects can be counted in the SSB masses in Eq.~(\ref{FTC}), whose values include also the loop corrections. 

\begin{figure}[ht!]
	\centering
	\includegraphics[scale=0.5]{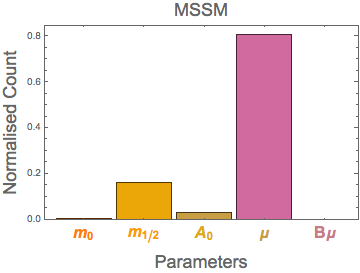}
	\includegraphics[scale=0.5]{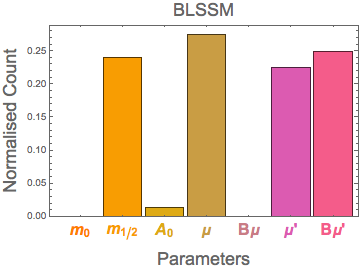}
	\caption{GUT-FT histogram for the MSSM (left) and BLSSM (right), showing contributions of the GUT-parameters.}
	\label{fig:histogram_BGFT}
\end{figure}

Indeed, if the required FT measure is quantified in terms of the GUT scale parameters, as done for the MSSM in \cite{Ellis:1985yc}, such as $m_{0},m_{1/2},A_{0},\mu, B\mu, \mu',B\mu'$, one can investigate what sector is the most effective in the required FT. Fig. \ref{fig:histogram_BGFT} displays the FT contributions of the fundamental parameters of  the MSSM and BLSSM. The dominating term in both cases is from the $\mu$ term, which is fixed (along with $B \mu$) by requiring EWSB. The next largest contribution to the FT measure arises from the gaugino sector, whose masses are parametrised via $m_{1/2}$. This can be understood with the heavy gluino mass bound \cite{gluino} and its large loop contribution to realise the 125 GeV Higgs boson. The BLSSM sector is also effective in the FT in terms of $\mu'$ and $B\mu'$. There is a very small dependence on $A_0$ as discussed previously, and approximately no dependence on $m_0$ or $B \mu$ in either case.

\begin{figure}
	\centering
	\includegraphics[scale=0.5]{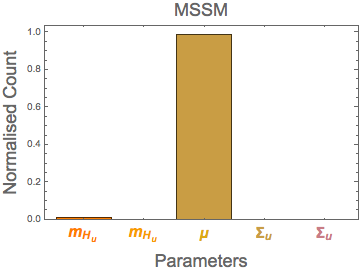} \includegraphics[scale=0.5]{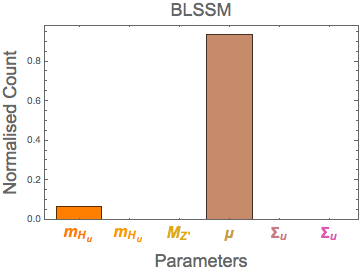}\caption{SUSY-FT histogram for the MSSM (left) and BLSSM (right), showing contribution of SUSY parameters. }
	\label{fig:EW-FT_histogram}
\end{figure}
Fig. \ref{fig:EW-FT_histogram} investigates which of the low-scale parameters are most responsible for the largest FT. Both the MSSM and BLSSM are dominated by the $\mu$'s FT, with a small contribution from $m_{H_u}$ and also a slight dependence on $M_{Z'}$ for the BLSSM. Considering this, what will affect the FT between the BLSSM and MSSM will be a combination of how large the factor $X$ is and the largeness of $\mu$ in both models. This value will not be identical, as there is an additional factor of $\frac{M_{Z'}(\tilde{g} Y)}{4g_{BL}}$ in the BLSSM minimisation equation.


\section{Results}
\label{sec:results}
We will now compare the FT obtained in the BLSSM and MSSM scenarios, for our two FT measures. We will begin by explaining the interval ranges of our data, then we will discuss the SUSY-scale  and GUT-scale FTs and which parameters are most responsible for their values. This will be done for both the BLSSM and MSSM, though the same parameters in both models are usually responsible for the largeness of FT. Then we will compare the GUT-FT and SUSY-FT for both the BLSSM and MSSM in the plane ($m_0$, $m_{1/2}$), as is commonly done.

The scan performed to obtain this data has been done by SPheno with all points being passed through {\texttt HiggsBounds} and {\texttt  HiggsSignals}. We have scanned over the range $[0,5]$ TeV in both $m_0$ and $m_{1/2}$, $\tan \beta$ in $[0 , 60]$, $A_0$ in $[-15, 15]$ TeV, which are common universal parameters for both the MSSM and the BLSSM, while for the BLSSM we also required $\tan \beta'$ in the interval $[0,2]$
with neutrino Yukawa couplings $Y^{(1,1)}$, $Y^{(2,2)}$, $Y^{(3,3)}$ in $[0,1]$. The $M_{Z'}$ value has been fixed to 4 TeV as discussed in Section~\ref{sec:colliderdm}. We will now compare the FT for both the MSSM and BLSSM, using both low- and high-scale parameters. 

\begin{figure}[t!]
	\centering
	\includegraphics[scale=0.56]{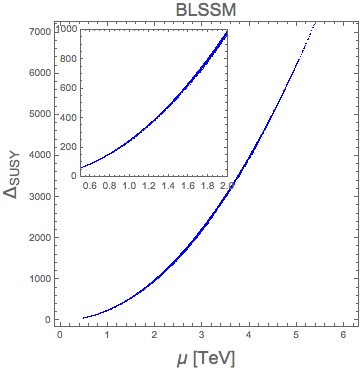}
	\caption{(a) SUSY-FT vs $\mu$. The very tight spread of points indicates $\mu$ is the dominant parameter responsible for SUSY-FT. }
	\label{fig:ft_susy_mu}
\end{figure}

We begin by presenting a measure of how the SUSY-FT  parameter varies with $\mu$ in the BLSSM. Fig. \ref{fig:ft_susy_mu} displays how the SUSY parameter FT varies with $\mu$. The FT measure is equal to the maximum contribution from any of the SUSY parameters, but here we see all data points centred on the curve. The tightness of our data shows that very rarely are the other ($m_{h_{u}}$, $m_{h_{d}}$, $\Sigma _u$, $\Sigma _d$) parameters ever responsible for the FT. This behaviour is expected, as one can see from the histogram plot of SUSY parameters, see Fig. \ref{fig:EW-FT_histogram}. The corresponding plot for the MSSM looks very similar and so is not shown. The behaviour is almost identical, as is expected from the MSSM version of the histogram discussed in section \ref{sec:fine_tuning}, whereby the $\mu$ parameter dominates the FT.

\begin{figure}[ht!]
	\centering
	\includegraphics[scale=0.56]{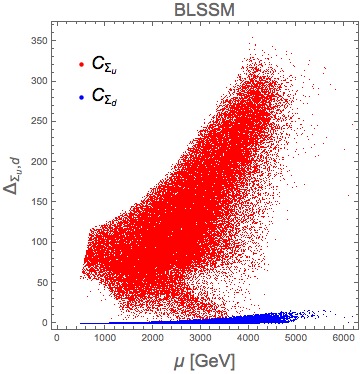}
	\caption{SUSY-FT for for one-loop tadpole corrections $C_{\Sigma_{u}}$ and $C_{\Sigma_{d}}$ for given values of $\mu$. Their contribution is never dominant and so loop corrections do not affect the SUSY-FT. }
	\label{fig:ft_susy_loop}
\end{figure}

Now, we turn our attention to considering loop contributions in the SUSY-scale  FT. By treating the loop factors as independent parameters which contribute to FT, we may observe their contributions. Fig. \ref{fig:ft_susy_loop} presents the contribution to FT from $\Sigma _u$ and $\Sigma _d$ whilst varying $\mu$. Immediately, one can compare the typical FT values with that of the overall FT as in Fig. \ref{fig:ft_susy_mu} and see that the loop contributions will never be the dominant contribution for the FT. There is some growth with $\mu$, but for any given value, the contribution from $\mu$ itself is $10$ times larger. Since only the maximum contribution of any $C_i$ parameter is taken, we find that treating the tadpole loop contributions as independent of the VEV  causes the one-loop FT to look much the same as at tree-level. Once again, this behaviour is mimicked in the MSSM, where the VEV independent tadpole loop corrections are also dwarfed by $\mu$'s FT.

\begin{figure}[ht!]
	\centering	
	\includegraphics[scale=0.56]{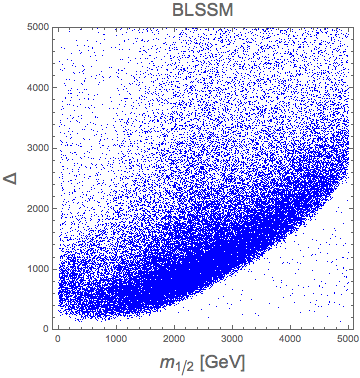}
	\caption{GUT-FT plotted against $m_{1/2}$. There is a strong, dependence for the GUT-FT with the $m_{1/2}$ parameter, although the wide upward spread indicates other parameters may also be the dominant FT contribution. }
	\label{fig:ft_gut_m12}
\end{figure}

Penultimately, before we turn to our final comparison of FT, we will discuss the dominant parameters in the GUT-FT sector. Fig. \ref{fig:ft_gut_m12} shows how the GUT-FT depends on $m_{1/2}$. There is a proportionality with $m_{1/2}$, favouring lower values for a better FT, but the points are not tightly constrained, unlike in SUSY-FT. The upward spread of points indicates that other parameters in addition to $m_{1/2}$ affect the FT. This is expected from the histogram in Fig. \ref{fig:histogram_BGFT}, where no one single parameter always determines FT, but rather a more even mix.

Finally, we will consider how the FT changes in the plane of ($m_0$, $m_{1/2}$). These two choices of parameters are selected since the universal scalar and gaugino masses are the two most important parameters. 
We colour the points with their FT values in four intervals, namely: red for FT > 5000, green for 1000 < FT < 5000, orange for 500 < FT < 1000 and blue (the least finely-tuned points) for FT < 500. The same set of points is used to compare the GUT-FT and the SUSY-FT (there is only a recolouring of these data points between left and right hand side) for the BLSSM and MSSM.
The overall picture is similar for all four cases and it is immediately clear that the FT is comparable between the BLSSM and the MSSM. There is a difference in the distribution of points between the MSSM and BLSSM, where there seem to be no viable points until $m_0 \sim 1$TeV in the latter. This is due to the requirement of a $Z'$ mass consistent with current constraints (see Section~\ref{sec:colliderdm}). Moreover, due to the tadpole equation given in Eq.~(\ref{BLmin}) relating $M_{Z'}$ to the soft-masses $m_{\eta_{1,2}}$, which are functions of $m_0$, notice that a larger $M_{Z'}$ leads to a larger $m_0$. 
All four graphs have a similar FT distribution, where a low $m_{1/2}$ is favoured and which manifests an approximate independence of $m_0$. Indeed, $m_{1/2}$ is mostly responsible for the FT rather than $m_0$ (see Fig.~\ref{fig:histogram_BGFT}). Since there is a little dependence on $m_0$, we expect to see an increasing FT as $m_{1/2}$ increases, as can be seen in all four cases.
When comparing the BLSSM and MSSM GUT-FT, the two pictures are very similar, with a slightly better FT in the MSSM, though the less fine tuned (blue) points appear about the same mass of $m_{1/2} \approx 2$ TeV.
This behaviour is very similar when comparing the SUSY-FT between BLSSM and MSSM, where the pictures (up to the distribution of points) are very similar, with a slight dependence on $m_0$, where larger values are favoured. 
Lastly, we compare the GUT-FT and SUSY-FT for each of the models. In the BLSSM we find a more concentrated region of less fine-tuned points at higher $m_0$. Both measures show a strong dependence on $m_{1/2}$. In the MSSM, we again find this dependence, but not the increase in density of less-finely tuned points as in the BLSSM.
To conclude the discussion on FT, we find that the overall FT is very comparable between the BLSSM and MSSM. Though the GUT-parameter measure is similar in both pictures, with the MSSM as slightly less finely tuned, the BLSSM has a larger density of less-finely-tuned points  when considering SUSY-parameters.

\clearpage
\begin{figure}[ht!]
	\newcommand{\size}{0.58}
	\subfigure[BLSSM GUT-FT.]{\includegraphics[scale=\size]{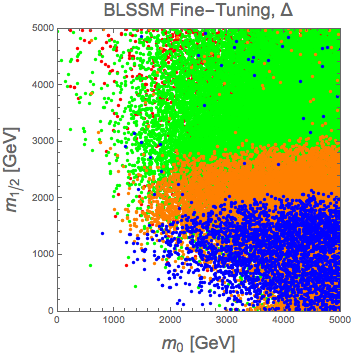} \label{fig:ftbg_blssm_m0_m12}}	
	\subfigure[BLSSM SUSY-FT. ]{\includegraphics[scale=\size]{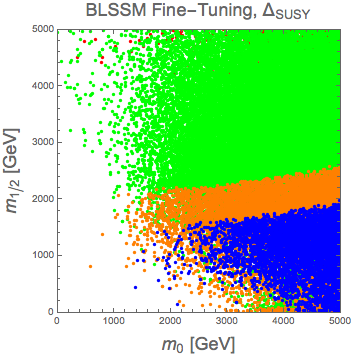}\label{fig:ftew_blssm_m0_m12}}

	\subfigure[MSSM GUT-FT.]{\includegraphics[scale=\size]{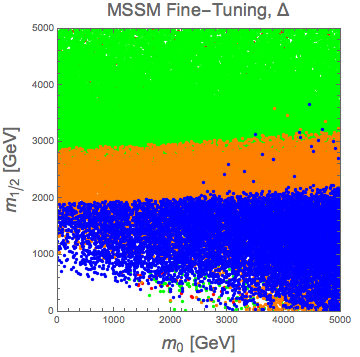}\label{fig:ftbg_mssm_m0_m12}}
	\subfigure[MSSM SUSY-FT. ]{\includegraphics[scale=\size]{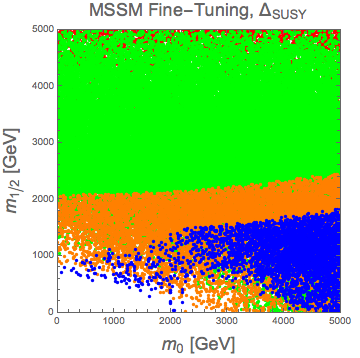}\label{fig:ftew_mssm_m0_m12}}
	\caption{Fine-tuning in the plane of unification of scalar, gaugino masses for BLSSM and MSSM for both GUT-parameters ($\Delta$) and EW parameters ($\Delta_{\rm EW}$). The FT is indicated by the colour of the dots: blue for FT $<$ 500; Orange for 500 $<$ FT $<$ 1000; Green for 1000 $<$ FT $<$ 500; and Red for FT $>$ 5000.}
	\label{fig:all_m0_m12}
\end{figure}

We now turn to considering the DM sectors of both models. We will see that once cosmological and direct detection bounds are imposed on the DM candidates, the BLSSM parameter space is far less constrained than the MSSM one, although at the cost of an increased GUT-FT.

\begin{figure}[ht!]
	\centering
	\includegraphics[scale=0.4]{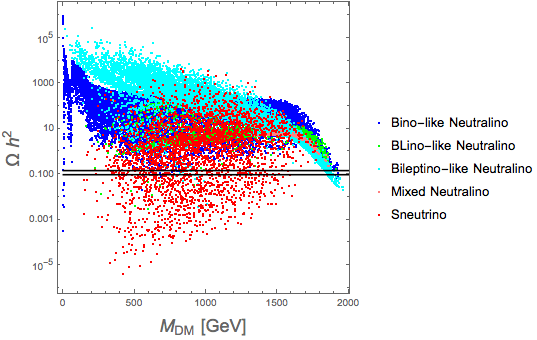}
	\includegraphics[scale=0.4]{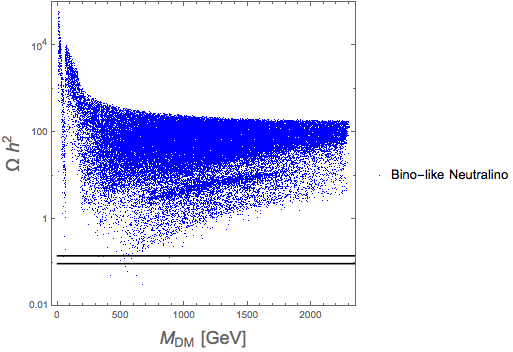}
	\caption{(a) Relic density vs LSP mass for the BLSSM.   
	         (b) Relic density vs LSP mass for the MSSM. In both plots the horizontal lines identify the $2\sigma$ region around the current central value of  $\Omega h^2$. }
	\label{fig:BLSSMvsMSSM-DM}
\end{figure}

For each generated spectrum, the LSP must comply with the cosmological and direct detection bounds of Section~\ref{sec:colliderdm}. The relic density in respect to the 
mass of the LSP ($M_{\rm DM}$) is plotted in Fig.~\ref{fig:BLSSMvsMSSM-DM}(a). 
The relic is overabundant for the large part of points surviving the screening from collider constraints. 
Without specifying initial conditions, as those 
igniting a favourable cohannihilation, our scan reveals multiple extended areas with relic densities close to zero.
Interestingly, the BLSSM successfully accommodates values within the allowed interval in Eq.~(\ref{PLANCK}), with all LSP species.
The corresponding distributions in Fig.~\ref{fig:BLSSMvsMSSM-DM}(a) have recognisable shapes, which point to different areas where a given LSP is more likely to cross
the experimentally allowed interval. Neutralinos may be found mostly, but not entirely, at large $M_{\rm DM}$ values. Sneutrinos appear in a cloud, with low relic density values around the centre of our mass span. The sneutrino option stands out as a very promising one, compensating its low rate of production as a LSP with a milder value of the relic with respect to the neutralino. 

The extended particle spectrum of the BLSSM yields a more varied nature
of the LSP, with more numerous combinations of DM annihilation diagrams, and can play a significant role in dramatically changing the response of the model to the cosmological data, in comparison to the much constrained MSSM.
This is well manifested by the relic density computed in the MSSM, as shown in Fig.~\ref{fig:BLSSMvsMSSM-DM}(b). From here, it is obvious how the BLSSM offers a variety of solutions to saturate the relic abundance compatible with the constraints, whether taken at $2\sigma$ from the central value measured by experiment or as an absolute
upper limit, precluded to the MSSM. In the former, different DM incarnations (Bino-, BLino-, Bileptino-like and mixed neutralino,
alongside the sneutrino) can comply with experimental evidence over a
$M_{\rm DM}$ interval which extends up to 2 TeV or so, while in the MSSM case solutions can only be found for much lighter LSP
masses and limitedly to one nature (the usual Bino-like neutralino).
Together with the limit on the cosmological relic produced at decoupling by the candidate DM particle, we challenge the constrained BLSSM against the negative
search for  Weakly Interactive Massive Particle (WIMP) nuclear recoils by the LUX experiment.
\begin{figure}[ht!]
	\centering
	\includegraphics[scale=0.45]{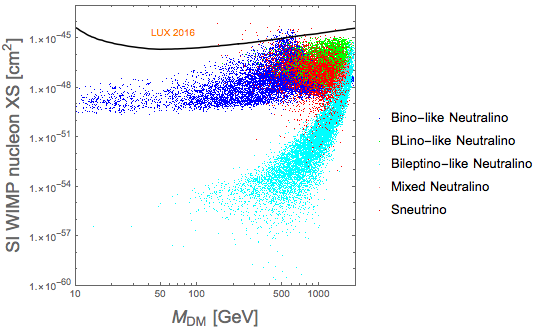}
	\caption{Spin-independent WIMP-nucleus scattering cross section generated in our scan against the upper bounds from 2016 run of the LUX experiment.}
	\label{LUX2016}
\end{figure}

The 2016 results of the LUX collaboration have seen the upper bound on the cross section decreasing by a factor of four in the three years of exposure. 
Such constraining analyses are still ongoing and will interestingly become a threat or a confirmation of the WIMP hypothesis in future years. 
From Fig.~\ref{LUX2016} we notice how the BLSSM with the parameter space investigated largely survives such tight limits. It is also worth stressing how the LUX bounds have started touching the BLSSM parameter space, so the next improvements of direct DM searches are  going to test the BLSSM very closely.

\section{Conclusions}
\label{sec:conclusions}

While several studies of the SUSY version of the $B-L$ model, BLSSM for short, exist for its  low energy phenomenology, predicting distinctive experimental signatures, 
very little had been said about the theoretical degree of FT required in this scenario in order to produce them or else to  escape current experimental constraints coming from
EWPOs, collider and cosmological data. We have done so in the first part of this paper, by adopting a suitable
FT measure amongst those available in literature and expressed it  in terms of the low energy spectra of the
MSSM and BLSSM as well as of the (high-scale) universal parameters of  the two models. The latter, 
for the MSSM, include: masses for scalars and gauginos, trilinear coupling, Higgsino mass and the quadratic soft SUSY term. In the BLSSM, we have all of these parameters plus two additional ones, the BLino mass and another quadratic soft SUSY term. The low and high energy spectra in the two SUSY scenarios can be related by RGEs,
that we have computed numerically at two-loop level.  

We have found that the level of FT required in the BLSSM is somewhat higher than in the MSSM when computed at the GUT scale in presence of all available experimental constraints, but those connected to DM searches, 
and this is primarily driven by the requirement of a large $Z'$ mass, of order 4 TeV or higher, which in turn corresponds to somewhat different acceptable values for the scalar and fermionic unification masses, which partially reflect in different low energy spectra potentially accessible at the LHC. However, when the FT is computed at the SUSY scale, the pull now originating from all available experimental constraints, chiefly the DM ones, destabilises the MSSM more than the BLSSM, as the latter appears more natural, well reflecting a much lower level of tension against data existing in the latter with respect to the former.

Furthermore, we have examined the response to the relic density constraints of  the non-minimal SUSY scenario,
 wherein the extra $B-L$
neutralinos (three extra neutral fermions: $U(1)_{B-L}$ gaugino
$\widetilde B'$ and two extra Higgsinos
$\widetilde{\eta}$) can be cold DM candidates.  
As well known, taking the lightest neutralino in
the MSSM as the sole possible DM  candidate implies severe constraints on the parameter space of this scenario.
Indeed, in the case of universal soft-breaking terms, the MSSM is
almost ruled out by combining  collider, astrophysics and rare
decay constraints. Therefore, it is important
to explore very well motivated extensions of the MSSM, such as the BLSSM, that provide new DM candidates that may account
for the relic density with no conflict with other phenomenological
constraints.

After an extensive study in this direction, we have concluded  that the extended particle spectrum of the BLSSM, in turn translating into a more varied nature of the LSP as well as  a
more numerous combination of DM annihilation diagrams,  can play a significant role in dramatically changing the 
ability of SUSY to adapt to  cosmological data, in comparison to the much constrained MSSM. In fact, the BLSSM offers a variety of solutions to the   relic abundance constraint, whether taken at $2\sigma$ from the central value measured by experiment or as an absolute
upper limit, unavailable in the MSSM: such as, alongside the usual Bino- (and possibly sneutrino), also BLino- and Bleptino-like as well as mixed neutralino
can comply with experimental evidence over a
$M_{\rm DM}$ interval which extends up to 2 TeV or so while in the MSSM case solutions can only be found for much lighter LSP masses ($\sim 500$ GeV) and limitedly to one nature (the  intimated  standard Bino-like neutralino).

\section*{Acknowledgements}
SM is supported in part through the NExT Institute.
The work of LDR has been supported by the ``Angelo Della Riccia'' foundation and the STFC/COFUND Rutherford International Fellowship scheme.
The work of CM is supported by the ``Angelo Della Riccia'' foundation and by the Centre of Excellence project No TK133 ``Dark Side of the Universe''.
The work of SK is partially supported by the STDF project 13858. All authors acknowledge support
from the grant H2020-MSCA-RISE-2014 n. 645722 (NonMinimalHiggs).

\newpage
\bibliographystyle{JHEP}

\begin{thebibliography}{10}

\bibitem{Abdallah:2015hza}
W.~Abdallah and S.~Khalil, \emph{{MSSM Dark Matter in Light of Higgs and LUX
  Results}}, \href{http://dx.doi.org/10.1155/2016/5687463}{\emph{Adv. High
  Energy Phys.} {\bf 2016} (2016) 5687463},
  [\href{http://arxiv.org/abs/1509.07031}{{\tt 1509.07031}}].

\bibitem{Khalil:2006yi}
S.~Khalil, \emph{{Low scale $B$ - L extension of the Standard Model at the
  LHC}}, \href{http://dx.doi.org/10.1088/0954-3899/35/5/055001}{\emph{J. Phys.}
  {\bf G35} (2008) 055001}, [\href{http://arxiv.org/abs/hep-ph/0611205}{{\tt
  hep-ph/0611205}}].

\bibitem{Basso:2008iv}
L.~Basso, A.~Belyaev, S.~Moretti and C.~H. Shepherd-Themistocleous,
  \emph{{Phenomenology of the minimal B-L extension of the Standard model: Z'
  and neutrinos}},
  \href{http://dx.doi.org/10.1103/PhysRevD.80.055030}{\emph{Phys. Rev.} {\bf
  D80} (2009) 055030}, [\href{http://arxiv.org/abs/0812.4313}{{\tt
  0812.4313}}].

\bibitem{Basso:2009gg}
L.~Basso, A.~Belyaev, S.~Moretti, G.~M. Pruna and C.~H.
  Shepherd-Themistocleous, \emph{{Phenomenology of the minimal B-L extension of
  the Standard Model}}, {\emph{PoS} {\bf EPS-HEP2009} (2009) 242},
  [\href{http://arxiv.org/abs/0909.3113}{{\tt 0909.3113}}].

\bibitem{Basso:2010yz}
L.~Basso, S.~Moretti and G.~M. Pruna, \emph{{Phenomenology of the minimal $B-L$
  extension of the Standard Model: the Higgs sector}},
  \href{http://dx.doi.org/10.1103/PhysRevD.83.055014}{\emph{Phys. Rev.} {\bf
  D83} (2011) 055014}, [\href{http://arxiv.org/abs/1011.2612}{{\tt
  1011.2612}}].

\bibitem{Basso:2010as}
L.~Basso, A.~Belyaev, S.~Moretti and G.~M. Pruna, \emph{{Higgs phenomenology in
  the minimal $B-L$ extension of the Standard Model at LHC}},
  \href{http://dx.doi.org/10.1088/1742-6596/259/1/012062}{\emph{J. Phys. Conf.
  Ser.} {\bf 259} (2010) 012062}, [\href{http://arxiv.org/abs/1009.6095}{{\tt
  1009.6095}}].

\bibitem{Majee:2010ar}
S.~K. Majee and N.~Sahu, \emph{{Dilepton Signal of a Type-II Seesaw at CERN
  LHC: Reveals a TeV Scale B-L Symmetry}},
  \href{http://dx.doi.org/10.1103/PhysRevD.82.053007}{\emph{Phys. Rev.} {\bf
  D82} (2010) 053007}, [\href{http://arxiv.org/abs/1004.0841}{{\tt
  1004.0841}}].

\bibitem{Li:2010rb}
T.~Li and W.~Chao, \emph{{Neutrino Masses, Dark Matter and B-L Symmetry at the
  LHC}}, \href{http://dx.doi.org/10.1016/j.nuclphysb.2010.10.004}{\emph{Nucl.
  Phys.} {\bf B843} (2011) 396--412},
  [\href{http://arxiv.org/abs/1004.0296}{{\tt 1004.0296}}].

\bibitem{Perez:2009mu}
P.~Fileviez~Perez, T.~Han and T.~Li, \emph{{Testability of Type I Seesaw at the
  CERN LHC: Revealing the Existence of the B-L Symmetry}},
  \href{http://dx.doi.org/10.1103/PhysRevD.80.073015}{\emph{Phys. Rev.} {\bf
  D80} (2009) 073015}, [\href{http://arxiv.org/abs/0907.4186}{{\tt
  0907.4186}}].

\bibitem{Emam:2007dy}
W.~Emam and S.~Khalil, \emph{{Higgs and Z-prime phenomenology in B-L extension
  of the standard model at LHC}},
  \href{http://dx.doi.org/10.1140/epjc/s10052-007-0411-7}{\emph{Eur. Phys. J.}
  {\bf C52} (2007) 625--633}, [\href{http://arxiv.org/abs/0704.1395}{{\tt
  0704.1395}}].

\bibitem{Khalil:2012gs}
S.~Khalil and S.~Moretti, \emph{{Heavy neutrinos, Z' and Higgs bosons at the
  LHC: new particles from an old symmetry}},
  \href{http://dx.doi.org/10.4236/jmp.2013.41002}{\emph{J. Mod. Phys.} {\bf 4}
  (2013) 7--10}, [\href{http://arxiv.org/abs/1207.1590}{{\tt 1207.1590}}].

\bibitem{Khalil:2013in}
S.~Khalil and S.~Moretti, \emph{{A simple symmetry as a guide toward new
  physics beyond the Standard Model}},
  \href{http://dx.doi.org/10.3389/fphy.2013.00010}{\emph{Front.in Phys.} {\bf
  1} (2013) 10}, [\href{http://arxiv.org/abs/1301.0144}{{\tt 1301.0144}}].

\bibitem{Khalil:2016lgy}
S.~Khalil, \emph{{Radiative symmetry breaking in supersymmetric $B-L$ models
  with an inverse seesaw mechanism}},
  \href{http://dx.doi.org/10.1103/PhysRevD.94.075003}{\emph{Phys. Rev.} {\bf
  D94} (2016) 075003}, [\href{http://arxiv.org/abs/1606.09292}{{\tt
  1606.09292}}].

\bibitem{Khalil:2007dr}
S.~Khalil and A.~Masiero, \emph{{Radiative B-L symmetry breaking in
  supersymmetric models}},
  \href{http://dx.doi.org/10.1016/j.physletb.2008.06.063}{\emph{Phys. Lett.}
  {\bf B665} (2008) 374--377}, [\href{http://arxiv.org/abs/0710.3525}{{\tt
  0710.3525}}].

\bibitem{FileviezPerez:2010ek}
P.~Fileviez~Perez and S.~Spinner, \emph{{The Fate of R-Parity}},
  \href{http://dx.doi.org/10.1103/PhysRevD.83.035004}{\emph{Phys. Rev.} {\bf
  D83} (2011) 035004}, [\href{http://arxiv.org/abs/1005.4930}{{\tt
  1005.4930}}].

\bibitem{CamargoMolina:2012hv}
J.~E. Camargo-Molina, B.~O'Leary, W.~Porod and F.~Staub, \emph{{The Stability
  Of R-Parity In Supersymmetric Models Extended By U(1)$_{B-L}$}},
  \href{http://dx.doi.org/10.1103/PhysRevD.88.015033}{\emph{Phys. Rev.} {\bf
  D88} (2013) 015033}, [\href{http://arxiv.org/abs/1212.4146}{{\tt
  1212.4146}}].

\bibitem{Kikuchi:2008xu}
T.~Kikuchi and T.~Kubo, \emph{{Radiative B-L symmetry breaking and the Z-prime
  mediated SUSY breaking}},
  \href{http://dx.doi.org/10.1016/j.physletb.2008.07.059}{\emph{Phys. Lett.}
  {\bf B666} (2008) 262--268}, [\href{http://arxiv.org/abs/0804.3933}{{\tt
  0804.3933}}].

\bibitem{Fonseca:2011vn}
R.~M. Fonseca, M.~Malinsky, W.~Porod and F.~Staub, \emph{{Running soft
  parameters in SUSY models with multiple U(1) gauge factors}},
  \href{http://dx.doi.org/10.1016/j.nuclphysb.2011.08.017}{\emph{Nucl. Phys.}
  {\bf B854} (2012) 28--53}, [\href{http://arxiv.org/abs/1107.2670}{{\tt
  1107.2670}}].

\bibitem{Elsayed:2011de}
A.~Elsayed, S.~Khalil and S.~Moretti, \emph{{Higgs Mass Corrections in the SUSY
  B-L Model with Inverse Seesaw}},
  \href{http://dx.doi.org/10.1016/j.physletb.2012.07.066}{\emph{Phys. Lett.}
  {\bf B715} (2012) 208--213}, [\href{http://arxiv.org/abs/1106.2130}{{\tt
  1106.2130}}].

\bibitem{Basso:2012tr}
L.~Basso and F.~Staub, \emph{{Enhancing $h \to \gamma \gamma$ with staus in
  SUSY models with extended gauge sector}},
  \href{http://dx.doi.org/10.1103/PhysRevD.87.015011}{\emph{Phys. Rev.} {\bf
  D87} (2013) 015011}, [\href{http://arxiv.org/abs/1210.7946}{{\tt
  1210.7946}}].

\bibitem{O'Leary:2011yq}
B.~O'Leary, W.~Porod and F.~Staub, \emph{{Mass spectrum of the minimal SUSY B-L
  model}}, \href{http://dx.doi.org/10.1007/JHEP05(2012)042}{\emph{JHEP} {\bf
  05} (2012) 042}, [\href{http://arxiv.org/abs/1112.4600}{{\tt 1112.4600}}].

\bibitem{Basso:2012ew}
L.~Basso, A.~Belyaev, D.~Chowdhury, M.~Hirsch, S.~Khalil, S.~Moretti et~al.,
  \emph{{Proposal for generalised Supersymmetry Les Houches Accord for see-saw
  models and PDG numbering scheme}},
  \href{http://dx.doi.org/10.1016/j.cpc.2012.11.004}{\emph{Comput. Phys.
  Commun.} {\bf 184} (2013) 698--719},
  [\href{http://arxiv.org/abs/1206.4563}{{\tt 1206.4563}}].

\bibitem{Elsayed:2012ec}
A.~Elsayed, S.~Khalil, S.~Moretti and A.~Moursy, \emph{{Right-handed
  sneutrino-antisneutrino oscillations in a TeV scale Supersymmetric B-L
  model}}, \href{http://dx.doi.org/10.1103/PhysRevD.87.053010}{\emph{Phys.
  Rev.} {\bf D87} (2013) 053010}, [\href{http://arxiv.org/abs/1211.0644}{{\tt
  1211.0644}}].

\bibitem{Khalil:2015naa}
S.~Khalil and S.~Moretti, \emph{{The $B-L$ Supersymmetric Standard Model with
  Inverse Seesaw at the Large Hadron Collider}},
  \href{http://arxiv.org/abs/1503.08162}{{\tt 1503.08162}}.

\bibitem{Abdallah:2014fra}
W.~Abdallah, S.~Khalil and S.~Moretti, \emph{{Double Higgs peak in the minimal
  SUSY B-L model}},
  \href{http://dx.doi.org/10.1103/PhysRevD.91.014001}{\emph{Phys. Rev.} {\bf
  D91} (2015) 014001}, [\href{http://arxiv.org/abs/1409.7837}{{\tt
  1409.7837}}].

\bibitem{Basso:2012gz}
L.~Basso, B.~O'Leary, W.~Porod and F.~Staub, \emph{{Dark matter scenarios in
  the minimal SUSY B-L model}},
  \href{http://dx.doi.org/10.1007/JHEP09(2012)054}{\emph{JHEP} {\bf 09} (2012)
  054}, [\href{http://arxiv.org/abs/1207.0507}{{\tt 1207.0507}}].

\bibitem{Abdallah:2015hma}
W.~Abdallah, J.~Fiaschi, S.~Khalil and S.~Moretti, \emph{{Z$'$$-$induced
  invisible right-handed sneutrino decays at the LHC}},
  \href{http://dx.doi.org/10.1103/PhysRevD.92.055029}{\emph{Phys. Rev.} {\bf
  D92} (2015) 055029}, [\href{http://arxiv.org/abs/1504.01761}{{\tt
  1504.01761}}].

\bibitem{Abdallah:2015uba}
W.~Abdallah, J.~Fiaschi, S.~Khalil and S.~Moretti, \emph{{Mono-jet, -photon and
  -Z signals of a supersymmetric (B - L) model at the Large Hadron Collider}},
  \href{http://dx.doi.org/10.1007/JHEP02(2016)157}{\emph{JHEP} {\bf 02} (2016)
  157}, [\href{http://arxiv.org/abs/1510.06475}{{\tt 1510.06475}}].

\bibitem{Hammad:2016trm}
A.~Hammad, S.~Khalil and S.~Moretti, \emph{{LHC signals of a B-L supersymmetric
  standard model CP -even Higgs boson}},
  \href{http://dx.doi.org/10.1103/PhysRevD.93.115035}{\emph{Phys. Rev.} {\bf
  D93} (2016) 115035}, [\href{http://arxiv.org/abs/1601.07934}{{\tt
  1601.07934}}].

\bibitem{Hammad:2015eca}
A.~Hammad, S.~Khalil and S.~Moretti, \emph{{Higgs boson decays into $\gamma
  \gamma$ and Z$\gamma$ in the MSSM and the B-L supersymmetric SM}},
  \href{http://dx.doi.org/10.1103/PhysRevD.92.095008}{\emph{Phys. Rev.} {\bf
  D92} (2015) 095008}, [\href{http://arxiv.org/abs/1503.05408}{{\tt
  1503.05408}}].

\bibitem{Barbieri:1998uv}
R.~Barbieri and A.~Strumia, \emph{{About the fine tuning price of LEP}},
  \href{http://dx.doi.org/10.1016/S0370-2693(98)00577-2}{\emph{Phys. Lett.}
  {\bf B433} (1998) 63--66}, [\href{http://arxiv.org/abs/hep-ph/9801353}{{\tt
  hep-ph/9801353}}].

\bibitem{Cacciapaglia:2006pk}
G.~Cacciapaglia, C.~Csaki, G.~Marandella and A.~Strumia, \emph{{The Minimal Set
  of Electroweak Precision Parameters}},
  \href{http://dx.doi.org/10.1103/PhysRevD.74.033011}{\emph{Phys. Rev.} {\bf
  D74} (2006) 033011}, [\href{http://arxiv.org/abs/hep-ph/0604111}{{\tt
  hep-ph/0604111}}].

\bibitem{Wendell:2010md}
{\scshape Super-Kamiokande} collaboration, R.~Wendell et~al.,
  \emph{{Atmospheric neutrino oscillation analysis with sub-leading effects in
  Super-Kamiokande I, II, and III}},
  \href{http://dx.doi.org/10.1103/PhysRevD.81.092004}{\emph{Phys. Rev.} {\bf
  D81} (2010) 092004}, [\href{http://arxiv.org/abs/1002.3471}{{\tt
  1002.3471}}].

\bibitem{Aulakh:1999cd}
C.~S. Aulakh, A.~Melfo, A.~Rasin and G.~Senjanovic, \emph{{Seesaw and
  supersymmetry or exact R-parity}},
  \href{http://dx.doi.org/10.1016/S0370-2693(99)00708-X}{\emph{Phys. Lett.}
  {\bf B459} (1999) 557--562}, [\href{http://arxiv.org/abs/hep-ph/9902409}{{\tt
  hep-ph/9902409}}].

\bibitem{Holdom:1985ag}
B.~Holdom, \emph{{Two U(1)'s and Epsilon Charge Shifts}},
  \href{http://dx.doi.org/10.1016/0370-2693(86)91377-8}{\emph{Phys. Lett.} {\bf
  B166} (1986) 196--198}.

\bibitem{Hinshaw:2012aka}
{\scshape WMAP} collaboration, G.~Hinshaw et~al., \emph{{Nine-Year Wilkinson
  Microwave Anisotropy Probe (WMAP) Observations: Cosmological Parameter
  Results}},
  \href{http://dx.doi.org/10.1088/0067-0049/208/2/19}{\emph{Astrophys. J.
  Suppl.} {\bf 208} (2013) 19}, [\href{http://arxiv.org/abs/1212.5226}{{\tt
  1212.5226}}].

\bibitem{Ade:2015xua}
{\scshape Planck} collaboration, P.~A.~R. Ade et~al., \emph{{Planck 2015
  results. XIII. Cosmological parameters}},
  \href{http://dx.doi.org/10.1051/0004-6361/201525830}{\emph{Astron.
  Astrophys.} {\bf 594} (2016) A13},
  [\href{http://arxiv.org/abs/1502.01589}{{\tt 1502.01589}}].

\bibitem{Khalil:2015wua}
S.~Khalil and C.~S. Un, \emph{{Muon Anomalous Magnetic Moment in SUSY B-L Model
  with Inverse Seesaw}},
  \href{http://dx.doi.org/10.1016/j.physletb.2016.10.035}{\emph{Phys. Lett.}
  {\bf B763} (2016) 164--168}, [\href{http://arxiv.org/abs/1509.05391}{{\tt
  1509.05391}}].

\bibitem{Coriano:2015sea}
C.~Coriano, L.~Delle~Rose and C.~Marzo, \emph{{Constraints on abelian
  extensions of the Standard Model from two-loop vacuum stability and
  $U(1)_{B-L}$}}, \href{http://dx.doi.org/10.1007/JHEP02(2016)135}{\emph{JHEP}
  {\bf 02} (2016) 135}, [\href{http://arxiv.org/abs/1510.02379}{{\tt
  1510.02379}}].

\bibitem{Staub:2013tta}
F.~Staub, \emph{{SARAH 4 : A tool for (not only SUSY) model builders}},
  \href{http://dx.doi.org/10.1016/j.cpc.2014.02.018}{\emph{Comput. Phys.
  Commun.} {\bf 185} (2014) 1773--1790},
  [\href{http://arxiv.org/abs/1309.7223}{{\tt 1309.7223}}].

\bibitem{Porod:2003um}
W.~Porod, \emph{{SPheno, a program for calculating supersymmetric spectra, SUSY
  particle decays and SUSY particle production at e+ e- colliders}},
  \href{http://dx.doi.org/10.1016/S0010-4655(03)00222-4}{\emph{Comput. Phys.
  Commun.} {\bf 153} (2003) 275--315},
  [\href{http://arxiv.org/abs/hep-ph/0301101}{{\tt hep-ph/0301101}}].

\bibitem{Accomando:2016sge}
E.~Accomando, C.~Coriano, L.~Delle~Rose, J.~Fiaschi, C.~Marzo and S.~Moretti,
  \emph{{Z', Higgses and heavy neutrinos in U(1)' models: from the LHC to the
  GUT scale}}, \href{http://dx.doi.org/10.1007/JHEP07(2016)086}{\emph{JHEP}
  {\bf 07} (2016) 086}, [\href{http://arxiv.org/abs/1605.02910}{{\tt
  1605.02910}}].

\bibitem{Khachatryan:2016zqb}
{\scshape CMS} collaboration, V.~Khachatryan et~al., \emph{{Search for narrow
  resonances in dilepton mass spectra in proton-proton collisions at $\sqrt{s}$
  = 13 TeV and combination with 8 TeV data}},
  \href{http://arxiv.org/abs/1609.05391}{{\tt 1609.05391}}.

\bibitem{Bechtle:2008jh}
P.~Bechtle, O.~Brein, S.~Heinemeyer, G.~Weiglein and K.~E. Williams,
  \emph{{HiggsBounds: Confronting Arbitrary Higgs Sectors with Exclusion Bounds
  from LEP and the Tevatron}},
  \href{http://dx.doi.org/10.1016/j.cpc.2009.09.003}{\emph{Comput.Phys.Commun.}
  {\bf 181} (2010) 138--167}, [\href{http://arxiv.org/abs/0811.4169}{{\tt
  0811.4169}}].

\bibitem{Bechtle:2011sb}
P.~Bechtle, O.~Brein, S.~Heinemeyer, G.~Weiglein and K.~E. Williams,
  \emph{{HiggsBounds 2.0.0: Confronting Neutral and Charged Higgs Sector
  Predictions with Exclusion Bounds from LEP and the Tevatron}},
  \href{http://dx.doi.org/10.1016/j.cpc.2011.07.015}{\emph{Comput.Phys.Commun.}
  {\bf 182} (2011) 2605--2631}, [\href{http://arxiv.org/abs/1102.1898}{{\tt
  1102.1898}}].

\bibitem{Bechtle:2013wla}
P.~Bechtle, O.~Brein, S.~Heinemeyer, O.~St{\aa}l, T.~Stefaniak et~al.,
  \emph{{HiggsBounds-4: Improved Tests of Extended Higgs Sectors against
  Exclusion Bounds from LEP, the Tevatron and the LHC}},
  \href{http://arxiv.org/abs/1311.0055}{{\tt 1311.0055}}.

\bibitem{Bechtle:2015pma}
P.~Bechtle, S.~Heinemeyer, O.~St{\aa}l, T.~Stefaniak and G.~Weiglein,
  \emph{{Applying Exclusion Likelihoods from LHC Searches to Extended Higgs
  Sectors}}, \href{http://dx.doi.org/10.1140/epjc/s10052-015-3650-z}{\emph{Eur.
  Phys. J.} {\bf C75} (2015) 421}, [\href{http://arxiv.org/abs/1507.06706}{{\tt
  1507.06706}}].

\bibitem{Bechtle:2013xfa}
P.~Bechtle, S.~Heinemeyer, O.~St{\aa}l, T.~Stefaniak and G.~Weiglein,
  \emph{{HiggsSignals: Confronting arbitrary Higgs sectors with measurements at
  the Tevatron and the LHC}},
  \href{http://dx.doi.org/10.1140/epjc/s10052-013-2711-4}{\emph{Eur. Phys. J.}
  {\bf C74} (2014) 2711}, [\href{http://arxiv.org/abs/1305.1933}{{\tt
  1305.1933}}].

\bibitem{Belanger:2006is}
G.~Belanger, F.~Boudjema, A.~Pukhov and A.~Semenov, \emph{{MicrOMEGAs 2.0: A
  Program to calculate the relic density of dark matter in a generic model}},
  \href{http://dx.doi.org/10.1016/j.cpc.2006.11.008}{\emph{Comput. Phys.
  Commun.} {\bf 176} (2007) 367--382},
  [\href{http://arxiv.org/abs/hep-ph/0607059}{{\tt hep-ph/0607059}}].

\bibitem{Belanger:2013oya}
G.~Belanger, F.~Boudjema, A.~Pukhov and A.~Semenov, \emph{{micrOMEGAs$_3$: A
  program for calculating dark matter observables}},
  \href{http://dx.doi.org/10.1016/j.cpc.2013.10.016}{\emph{Comput. Phys.
  Commun.} {\bf 185} (2014) 960--985},
  [\href{http://arxiv.org/abs/1305.0237}{{\tt 1305.0237}}].

\bibitem{Akerib:2016lao}
{\scshape LUX} collaboration, D.~S. Akerib et~al., \emph{{Results on the
  Spin-Dependent Scattering of Weakly Interacting Massive Particles on Nucleons
  from the Run 3 Data of the LUX Experiment}},
  \href{http://dx.doi.org/10.1103/PhysRevLett.116.161302}{\emph{Phys. Rev.
  Lett.} {\bf 116} (2016) 161302}, [\href{http://arxiv.org/abs/1602.03489}{{\tt
  1602.03489}}].

\bibitem{Akerib:2016vxi}
D.~S. Akerib et~al., \emph{{Results from a search for dark matter in the
  complete LUX exposure}},  \href{http://arxiv.org/abs/1608.07648}{{\tt
  1608.07648}}.

\bibitem{Falk:1994es}
T.~Falk, K.~A. Olive and M.~Srednicki, \emph{{Heavy sneutrinos as dark
  matter}}, \href{http://dx.doi.org/10.1016/0370-2693(94)90639-4}{\emph{Phys.
  Lett.} {\bf B339} (1994) 248--251},
  [\href{http://arxiv.org/abs/hep-ph/9409270}{{\tt hep-ph/9409270}}].

\bibitem{Anderson:1994dz}
G.~W. Anderson and D.~J. Castano, \emph{{Measures of fine tuning}},
  \href{http://dx.doi.org/10.1016/0370-2693(95)00051-L}{\emph{Phys. Lett.} {\bf
  B347} (1995) 300--308}, [\href{http://arxiv.org/abs/hep-ph/9409419}{{\tt
  hep-ph/9409419}}].

\bibitem{Anderson:1994tr}
G.~W. Anderson and D.~J. Castano, \emph{{Naturalness and superpartner masses or
  when to give up on weak scale supersymmetry}},
  \href{http://dx.doi.org/10.1103/PhysRevD.52.1693}{\emph{Phys. Rev.} {\bf D52}
  (1995) 1693--1700}, [\href{http://arxiv.org/abs/hep-ph/9412322}{{\tt
  hep-ph/9412322}}].

\bibitem{Anderson:1995cp}
G.~W. Anderson and D.~J. Castano, \emph{{Challenging weak scale supersymmetry
  at colliders}}, \href{http://dx.doi.org/10.1103/PhysRevD.53.2403}{\emph{Phys.
  Rev.} {\bf D53} (1996) 2403--2410},
  [\href{http://arxiv.org/abs/hep-ph/9509212}{{\tt hep-ph/9509212}}].

\bibitem{Anderson:1996ew}
G.~W. Anderson, D.~J. Castano and A.~Riotto, \emph{{Naturalness lowers the
  upper bound on the lightest Higgs boson mass in supersymmetry}},
  \href{http://dx.doi.org/10.1103/PhysRevD.55.2950}{\emph{Phys. Rev.} {\bf D55}
  (1997) 2950--2954}, [\href{http://arxiv.org/abs/hep-ph/9609463}{{\tt
  hep-ph/9609463}}].

\bibitem{Ciafaloni:1996zh}
P.~Ciafaloni and A.~Strumia, \emph{{Naturalness upper bounds on gauge mediated
  soft terms}},
  \href{http://dx.doi.org/10.1016/S0550-3213(97)00138-7}{\emph{Nucl. Phys.}
  {\bf B494} (1997) 41--53}, [\href{http://arxiv.org/abs/hep-ph/9611204}{{\tt
  hep-ph/9611204}}].

\bibitem{Chan:1997bi}
K.~L. Chan, U.~Chattopadhyay and P.~Nath, \emph{{Naturalness, weak scale
  supersymmetry and the prospect for the observation of supersymmetry at the
  Tevatron and at the CERN LHC}},
  \href{http://dx.doi.org/10.1103/PhysRevD.58.096004}{\emph{Phys. Rev.} {\bf
  D58} (1998) 096004}, [\href{http://arxiv.org/abs/hep-ph/9710473}{{\tt
  hep-ph/9710473}}].

\bibitem{Giusti:1998gz}
L.~Giusti, A.~Romanino and A.~Strumia, \emph{{Natural ranges of supersymmetric
  signals}}, \href{http://dx.doi.org/10.1016/S0550-3213(99)00153-4}{\emph{Nucl.
  Phys.} {\bf B550} (1999) 3--31},
  [\href{http://arxiv.org/abs/hep-ph/9811386}{{\tt hep-ph/9811386}}].

\bibitem{Casas:2003jx}
J.~A. Casas, J.~R. Espinosa and I.~Hidalgo, \emph{{The MSSM fine tuning
  problem: A Way out}},
  \href{http://dx.doi.org/10.1088/1126-6708/2004/01/008}{\emph{JHEP} {\bf 01}
  (2004) 008}, [\href{http://arxiv.org/abs/hep-ph/0310137}{{\tt
  hep-ph/0310137}}].

\bibitem{Casas:2004uu}
J.~A. Casas, J.~R. Espinosa and I.~Hidalgo, \emph{{A Relief to the
  supersymmetric fine tuning problem}},  in \emph{{String phenomenology.
  Proceedings, 2nd International Conference, Durham, UK, July 29-August 4,
  2003}}, pp.~76--85, 2004.
\newblock \href{http://arxiv.org/abs/hep-ph/0402017}{{\tt hep-ph/0402017}}.

\bibitem{Casas:2004gh}
J.~A. Casas, J.~R. Espinosa and I.~Hidalgo, \emph{{Implications for new physics
  from fine-tuning arguments. 1. Application to SUSY and seesaw cases}},
  \href{http://dx.doi.org/10.1088/1126-6708/2004/11/057}{\emph{JHEP} {\bf 11}
  (2004) 057}, [\href{http://arxiv.org/abs/hep-ph/0410298}{{\tt
  hep-ph/0410298}}].

\bibitem{Casas:2006bd}
J.~A. Casas, J.~R. Espinosa and I.~Hidalgo, \emph{{Expectations for LHC from
  naturalness: modified versus SM Higgs sector}},
  \href{http://dx.doi.org/10.1016/j.nuclphysb.2007.04.024}{\emph{Nucl. Phys.}
  {\bf B777} (2007) 226--252}, [\href{http://arxiv.org/abs/hep-ph/0607279}{{\tt
  hep-ph/0607279}}].

\bibitem{Kitano:2005wc}
R.~Kitano and Y.~Nomura, \emph{{A Solution to the supersymmetric fine-tuning
  problem within the MSSM}},
  \href{http://dx.doi.org/10.1016/j.physletb.2005.10.003}{\emph{Phys. Lett.}
  {\bf B631} (2005) 58--67}, [\href{http://arxiv.org/abs/hep-ph/0509039}{{\tt
  hep-ph/0509039}}].

\bibitem{Athron:2007ry}
P.~Athron and D.~J. Miller, \emph{{A New Measure of Fine Tuning}},
  \href{http://dx.doi.org/10.1103/PhysRevD.76.075010}{\emph{Phys. Rev.} {\bf
  D76} (2007) 075010}, [\href{http://arxiv.org/abs/0705.2241}{{\tt
  0705.2241}}].

\bibitem{Baer:2012up}
H.~Baer, V.~Barger, P.~Huang, A.~Mustafayev and X.~Tata, \emph{{Radiative
  natural SUSY with a 125 GeV Higgs boson}},
  \href{http://dx.doi.org/10.1103/PhysRevLett.109.161802}{\emph{Phys. Rev.
  Lett.} {\bf 109} (2012) 161802}, [\href{http://arxiv.org/abs/1207.3343}{{\tt
  1207.3343}}].

\bibitem{Ellis:1985yc}
J.~R. Ellis, K.~Enqvist, D.~V. Nanopoulos and F.~Zwirner, \emph{{Aspects of the
  Superunification of Strong, Electroweak and Gravitational Interactions}},
  \href{http://dx.doi.org/10.1016/0550-3213(86)90015-5}{\emph{Nucl. Phys.} {\bf
  B276} (1986) 14--70}.

\bibitem{Barbieri:1987fn}
R.~Barbieri and G.~F. Giudice, \emph{{Upper Bounds on Supersymmetric Particle
  Masses}}, \href{http://dx.doi.org/10.1016/0550-3213(88)90171-X}{\emph{Nucl.
  Phys.} {\bf B306} (1988) 63--76}.

\bibitem{Ross:2017kjc}
G.~G. Ross, K.~Schmidt-Hoberg and F.~Staub, \emph{{Revisiting fine-tuning in
  the MSSM}},  \href{http://arxiv.org/abs/1701.03480}{{\tt 1701.03480}}.

\bibitem{gluino}
T.~A. collaboration, \emph{{Search for pair-production of gluinos decaying via
  stop and sbottom in events with $b$-jets and large missing transverse
  momentum in $\sqrt{s}=13$ TeV $pp$ collisions with the ATLAS detector}}, .

\end{thebibliography}

\providecommand{\href}[2]{#2}\begingroup\raggedright\endgroup

\end{document}